\documentclass[twocolumn]{aastex63}

\accepted{to ApJ \today}

\usepackage{graphicx}
\usepackage{dcolumn}
\usepackage{bm}
\usepackage{mhchem}
\usepackage{longtable}
\usepackage{supertabular}

\usepackage{textpos}
\usepackage{hyperref}
\usepackage{amsmath}
\usepackage{wasysym}

\newcommand{\beginsupplement}{%
        \setcounter{table}{0}
        \renewcommand{\thetable}{A\arabic{table}}%
        \setcounter{figure}{0}
        \renewcommand{\thefigure}{A\arabic{figure}}%
        \setcounter{equation}{0}
        \renewcommand{\theequation}{A\arabic{equation}}%
     }

\shorttitle{HCN production in Titan's Atmosphere}
\shortauthors{Pearce et al.}

\begin{document}

\title{HCN production in Titan's Atmosphere: Coupling quantum chemistry and disequilibrium atmospheric modeling}

\correspondingauthor{Ben K. D. Pearce}
\email{pearcbe@mcmaster.ca}

\author{Ben K. D. Pearce}
\affiliation{Origins Institute and Department of Physics and Astronomy, McMaster University, ABB 241, 1280 Main St, Hamilton, ON, L8S 4M1, Canada}

\author{Karan Molaverdikhani}
\affiliation{Planet and Star Formation Department, Max Planck Institute for Astronomy, 69117 Heidelberg, Germany}

\author{Ralph E. Pudritz}
\affiliation{Origins Institute and Department of Physics and Astronomy, McMaster University, ABB 241, 1280 Main St, Hamilton, ON, L8S 4M1, Canada}

\author{Thomas Henning}
\affiliation{Planet and Star Formation Department, Max Planck Institute for Astronomy, 69117 Heidelberg, Germany}

\author{Eric H{\'e}brard}
\affiliation{Astrophysics Group, University of Exeter, Exeter, Devon EX4 4QL, UK}

\begin{abstract}

Hydrogen cyanide (HCN) is a critical reactive source of nitrogen for building key biomolecules relevant for the origin of life. Still, many HCN reactions remain uncharacterized by experiments and theory, and the complete picture of HCN production in planetary atmospheres is not fully understood. To improve this situation, we develop a novel technique making use of computational quantum chemistry, experimental data, and atmospheric numerical simulations. First, we use quantum chemistry simulations to explore the entire field of possible reactions for a list of primary species in \ce{N2}-, \ce{CH4}-, and \ce{H2}-dominated atmospheres. In this process, we discover 33 new reactions with no previously known rate coefficients. 
From here, we develop a consistent reduced atmospheric hybrid chemical network (CRAHCN) containing experimental values when available, and our calculated rate coefficients otherwise. Next, we couple CRAHCN to a 1D chemical kinetic model (ChemKM) to compute the HCN abundance as a function of atmospheric depth on Titan. 
Our simulated atmospheric HCN profile agrees very well with the Cassini observations. CRAHCN contains 104 reactions however nearly all of the simulated atmospheric HCN profile can be obtained using a scaled down network of only 19 dominant reactions. From here, we form a complete picture of HCN chemistry in Titan's atmosphere, from the dissociation of the main atmospheric species, down to the direct production of HCN along 4 major channels. One of these channels was first discovered and characterized in \citet{Reference598} and this work.
\end{abstract}


\section{Introduction}

Hydrogen cyanide (HCN) is a fundamental molecule in the origins of life. Both nucleobases, the building blocks of DNA/RNA, and amino acids, the building blocks of proteins form in HCN reactions \citep{1961Natur.191.1193O,Reference600}. Consequently, a terrestrial atmosphere rich in HCN may be a distinct feature of what we term a \emph{biogenic planet}, i.e. a planet capable of producing key biomolecules without requiring exogenous sources (e.g. meteorites).

In atmospheres, HCN generally forms out of the reactive radicals left over from methane (\ce{CH4}) and nitrogen (\ce{N2}) dissociation \citep{2007AsNow..22e..76R,Reference598,2017JGRE..122..432H}. These radicals are \ce{^4N}, \ce{^2N}, \ce{CH3}, \ce{^3CH2}, \ce{^1CH2}, \ce{CH}, \ce{H2} and \ce{H}, where the leading superscripts signify the singlet, doublet, triplet, and quartet spin states \citep{2017ApJ...850...48S,Reference586}. There are various energy sources capable of dissociating \ce{CH4} and \ce{N2} in an atmosphere, including ultraviolet (UV) light, galactic cosmic rays (GCRs), and lightning.

The most HCN-rich atmosphere in the Solar System belongs to Saturn's moon Titan. From 2004--2009, 4 instruments aboard the Cassini spacecraft measured the HCN molar mixing ratios\footnote{Molar mixing ratios are the molar abundances of species divided by that of the entire atmospheric composition.} in Titan's atmosphere to be $\sim$0.1--10ppm (parts-per-million) in the lower atmosphere ($<$600 km), and $\sim$0.1--5$\permil$ (parts-per-thousand) in the upper atmosphere ($>$ 700 km) \citep{2010Icar..205..559V,2011Icar..214..584A,2011Icar..216..507K,2009PSS...57.1895M}. Titan has a surface temperature of $\sim$94 K, and an atmospheric composition of approximately 1.5 bars of N$_2$ ($\sim$94.2\%), CH$_4$ ($\sim$5.7\%) and H$_2$ ($\sim$0.1\%) with relatively low abundances of oxygen species (\ce{CO}: 40--50 ppm, \ce{CO2}: 10--20 ppb, \ce{H2O}: 0.5--8 ppb) \citep{Reference143,2017JGRE..122..432H}. UV light and GCRs are responsible for dissociating \ce{N2} and \ce{CH4} to produce radical species in Titan's upper and lower atmospheres, respectively \citep{2019Icar..324..120V,2011AA...529A.143G,2009AA...506..955G}. Numerical simulations of \ce{N2}-rich exoplanet atmospheres suggest Titan's high atmospheric HCN composition is caused by its high atmospheric C/O ratio ($\gg$ 1) \citep{2019Icar..329..124R}.

HCN has also been detected in the atmospheres of Pluto and Neptune with concentrations of $\sim$40 ppm and $\sim$1 ppb (parts-per-billion), respectively \citep{2017Icar..286..289L,1993ApJ...406..285M}. Observations have also been used to put an upper bound of 0.1 ppb on the HCN concentration in Uranus's atmosphere \citep{1993ApJ...406..285M}. Finally, HCN has been tentatively detected in the exoplanet atmospheres of 55 Cancri e \citep{2016ApJ...820...99T} and WASP-63b \citep{MacDonald_2017}, and may have been present in the early Earth atmosphere prior to the origin of life \citep{2007AsNow..22e..76R}.

Given the abundance of HCN in Titan's atmosphere, and the availability of the Cassini data, Titan is the perfect testbed for validating theoretical chemical networks for HCN production in atmospheres. In the past, large-scale networks containing 800--3000+ reactions have been paired with 1D chemical kinetic codes to calculate the HCN profile as well as other chemical profiles in Titan's atmosphere \citep{2019Icar..324..120V,2016ApJ...829...79W,2015Icar..247..218L,2014PSS..104...48L,2012AA...541A..21H,2009Icar..201..226K,2008PSS...56...67L,2007PSS...55.1470H}. Past simulations provide a reasonable agreement with the Cassini atmospheric HCN measurements, and agree on the importance of \ce{H2CN + H -> HCN + H2} as a pathway for HCN production, and HCN photolysis (\ce{HCN + $h \nu$ -> CN + H}) as a destruction process \citep{2012AA...541A..21H,2015Icar..247..218L,2016ApJ...829...79W,2019Icar..324..120V}. Rate coefficients in these networks are typically gathered from a variety of sources with differing accuracies (e.g. experiments, theoretical simulations, similar reactions, thermodynamics). Quantum chemistry methods are also occasionally used to introduce new reaction rate coefficients (e.g. \citet{2012AA...541A..21H,2015Icar..247..218L,2019Icar..324..120V}).

Up until this point, there have been gaps in the HCN chemical data, preventing simulations from obtaining a complete picture of HCN production and destruction in Titan's atmosphere. There is a particular absence of rate coefficient data for reactions involving excited species such as doublet nitrogen atoms (\ce{^2N}) and singlet methylene (\ce{^1CH2}): two species that are directly produced from ultraviolet (UV) and galactic cosmic ray (GCR) dissociation in Titan's upper and lower atmosphere, respectively \citep{2019Icar..324..120V}. 


We take the next step in simulating HCN chemistry in Titan's atmosphere, by finding, calculating, and including all the missing reactions relevant to the production and destruction of HCN in this environment and any direct competing reactions. Furthermore, given the complexity and computational cost of analyzing large-scale networks, we take a different approach from past Titan simulations by building and implementing a reduced network. What we have discovered, is that HCN chemistry can really be understood with a highly reduced set of chemical reactions, and that this approach is invaluable to obtaining physical insights from the results. A reduced network approach has been taken in the past for modeling warm to hot hydrogen-dominated atmospheres, and has similarly been found to conserve most of the information and insight into the dominant production and destruction pathways of observable species \citep{venot2019reduced}. Moreover, shifting the focus to the development of smaller chemical networks that are less computationally demanding would make them implementable in 3D atmosphere models (e.g., \citet{Reference2071}), which are the only ones that are able to reproduce seasonal effects.

To summarize our method, we develop a ``bottom up'' theoretical approach to analyze the main formation and destruction channels of HCN in Titan's atmosphere. We start by using computational quantum chemistry simulations to scan all possible reactions between a small set of primary species. The primary species are the main reactive constituents of Titan's atmosphere (\ce{CH4, N2, H2, HCN}), their dissociation products, and a few key intermediates (see Table~\ref{Table1} for list of primary species). We then use canonical variational transition state theory \citep{Reference534,Reference598} and Rice--Ramsperger--Kassel--Marcus/master equation (RRKM/ME) theory \citep{Reference2058} to calculate the rate coefficients for all these reactions. Many of these reactions are discovered here for the first time. We validate these calculations by comparing with experimental values in the 32\% of cases that are available.

From this, we combine our calculated rate coefficients with available experimental values to obtain a consistent and complete chemical network for the reduced set of primary species. We call this network CRAHCN (Consistent Reduced Atmospheric Hybrid Chemical Network). We then couple CRAHCN with a 1D chemical kinetic model (ChemKM) \citep{Molaverdikhani_2019} in order to simulate the production and destruction of HCN in Titan's atmosphere. 

This approach is beneficial in that it allows us to a) accept or reject previously reported reactions, b) discover previously unknown and potentially very important reactions, and c) develop a very fast, accurate, and consistent code to compute HCN chemistry in atmospheres. 

We began this effort in \citet{Reference598}, where we developed a feasible and accurate method to calculate a small network of 41 unique reaction rate coefficients that are directly involved with or in competition with the production of HCN in atmospheres. We focused mainly on validating the method using reactions previously studied by theory or experiment; However 15 of the calculated rate coefficients in that work had no previously known values. 

In this work, we expand the network to include CN as a primary reacting species. We also explore more deeply into the unknown territory of HCN chemistry by simulating \emph{all} the possible efficient interactions between the primary molecular species in the network. In this expansion, we also include three-body reactions, where an atmospheric molecule collisionally deexcites a vibrationally excited intermediate. Next, we modify some of the reactions from \citet{Reference598} due to new knowledge about vibrationally excited and unstable intermediates (see theoretical case studies in Appendix materials). Finally, we include two experimental spin-forbidden collisionally induced intersystem crossing reactions \citep{Reference2050,Reference552} whose rate coefficients cannot be calculated using our theoretical method. We direct the reader to the Appendix Tables for the CRAHCN rate coefficient data.

In this process, we discover 33 brand new reactions, the majority of which are based on \ce{H2CN}, \ce{CN}, \ce{CH}, and electronically excited molecules (\ce{^2N} and \ce{^1CH2}). Ultimately, we finish with a consistent reduced network containing 104 reactions, which is complete for the 14 primary species in this work. In the end, we discover that only 19 reactions are at the heart of HCN production and destruction in Titan's atmosphere.


This paper proceeds as follows: in Section~\ref{methods} we describe the theoretical and computational quantum chemistry methods used to calculate the reaction rate coefficients in CRAHCN, and we outline the model parameters for our atmospheric numerical simulations of Titan. Then, in Section~\ref{results1} we describe the results of the rate coefficient calculations, and their conformance to any experimentally measured values. We also perform a methods comparison to compare the accuracies of our chosen computational quantum chemistry method and two other widely used methods. Next, in Section~\ref{results2} we analyze the results of our four numerical models of Titan's atmosphere: our fiducial model, a model with only 19 dominant reactions, a model with no GCRs, and a model with a different input for eddy diffusion. We then compare our results to the three most recent Titan models in the literature \citep{2019Icar..324..120V,2016ApJ...829...79W,2015Icar..247..218L}. In this section, we also describe the two sensitivity analyses which allow us to identify the dominant pathways to HCN production and destruction on Titan. Sensitivity analyses involve running simulations where reactions from the network are excluded I) one at a time, and II) multiple at a time. Then, in Section~\ref{discuss} we present a step-by-step guide to the production of HCN in Titan's atmosphere, and we discuss how CRAHCN can be used for other atmospheric models. Finally, we summarize our main conclusions in Section~\ref{conclusions}.
 
The Appendix materials contains a) three tables comparing calculated rate coefficients at 298 K with experimental values, b) two tables containing the CRAHCN reaction rate coefficients, expressed as their Lindemann and Arrhenius parameters for temperatures from 50--400 K, c) any experimental data for the reactions calculated in this work, d) a breakdown of the calculations for some of the non-standard reactions in CRAHCN, and e) the raw computational quantum chemistry data used for rate coefficient calculations.


\begin{table}[ht!]
\centering
\caption{List of primary molecular species in this network and their spin states. \label{Table1}} 
\begin{tabular}{lcc}
\\
\multicolumn{1}{l}{Species} & 
\multicolumn{1}{c}{Spin state} & 
\multicolumn{1}{c}{Ground/Excited state}\\ \hline \\[-2mm]
HCN & singlet & ground \\
H$_2$CN & doublet & ground \\
N$_2$ & singlet & ground \\
CN & doublet & ground \\
$^2$N & doublet & excited \\
$^4$N & quartet & ground \\
CH$_4$ & singlet & ground \\
CH$_3$ & doublet & ground \\
$^1$CH$_2$ & singlet & excited \\
$^3$CH$_2$ & triplet & ground \\
CH & doublet & ground \\
H$_2$ & singlet & ground \\
H & doublet & ground \\
NH & triplet & ground \\
\hline
\end{tabular}
\end{table}

\section{Methods}\label{methods}

Our atmospheric model can be roughly divided into two components, a) the chemical network, which is the collection of reactions and their experimental or calculated rate coefficients, and b) the chemical kinetic code, which handles radiative transfer, molecular and eddy diffusion, molecular influx and escape, photodissociation, and GCR dissociation.

\subsection{Rate Coefficient Calculations}

There are 104 reactions in our reduced network, representing all efficient reactions between the 14 primary species in our network (see Table~\ref{Table1}). Our strategy for building up the network is as follows.

Firstly, if experimental data is available, we use it in this network. This accounts for 42 of the 104 reactions.

Second, we use a standard, fast, and accurate computational quantum chemistry method combined with standard theoretical methods to compute all possible rate coefficients and compare our results with any experimental values. This computational method is the Becke-Half-and-Half-Lee-Yang-Parr\footnote{Hartree-Fock (HF) methods tend to over-estimate energy barriers, and density functional theory (DFT) methods tend to under-estimate energy barriers. BHandHLYP offers a reasonable solution by using 50\% Hartree-Fock and 50\% density functional theory for the exchange energy calculation.} (BHandHLYP) density functional and the augmented correlation-consistent polarized valence double-$\zeta$ (aug-cc-pVDZ) basis set\footnote{The basis set is the defined space for the problem, in our case it represents the atomic orbitals. The aug-cc-pVDZ basis set includes all atomic orbitals within the electron shell that is 1 above the atom's valence shell.} \citep{Reference594,Reference595,Dunning1989,KendallDunning1992,WoonDunning1993}. We call this method BH/d for short. We show that there is a good agreement between experimental and calculated rate coefficients when using this computational method, with the majority of calculations (64\%) landing within a factor of 2 of experimental values and all values landing within about an order of magnitude of experimental values. 

Typical uncertainties for rate coefficients---assigned in large-scale experimental data evaluations---range from a factor of 2 to an order of magnitude \citep{Reference451,Reference509}. As examples in our network, \citet{Reference451} assign factor of 2--3 uncertainties to the rate coefficients of \ce{H + H + M -> H2 + M} and \ce{CH3 + H + M -> CH4 + M}, and order of magnitude uncertainties to the rate coefficients of \ce{CH4 + CH -> products} and \ce{CN + ^4N -> N2 + C}.

Third, we compare the accuracy of our BH/d calculations with A) a second DFT method, and B) an {\it ab inito} method. The DFT method we use for comparison is the fairly recently developed asymptotically corrected $\omega$B97XD functional \citep{ChaiGordon2008}. We again use the aug-cc-pVDZ basis set, and call this method $\omega$B/d for short. The {\it ab inito} method is coupled-cluster singles and doubles \footnote{The major benefit of coupled-cluster theory over DFT tends to come into play for systems with strong electron correlation effects. Coupled-cluster methods are able to describe the quantum many-body effects of the electronic wave function at a computational cost is significantly more expensive than DFT. Coupled-cluster methods are size-extensive, and thus provide a correct scaling for the correlation energy with respect to the number of electrons \citep{Reference2066,Reference2067,Reference2068}.} (CCSD) with the aug-cc-pVTZ basis set \citep{Reference2062}. For convenience, we designate this as CC/t. We will show below that $\omega$B/d and CC/t do not necessarily improve the accuracy of our calculations in terms of agreement with experimental rate coefficients. In 8 out of 12 cases, BH/d gives the best, or equal to the best agreement with experiment in comparison with the other two methods. The other two methods give the best, or equal to the best agreement in 7 out of 12 and 6 out of 12 cases, respectively. We summarize these results in detail in Section~\ref{methodcompare}. For recent detailed reviews of DFT and coupled-cluster theory, we refer the reader to \citet{Reference2070}, and \citet{Reference2069}, respectively.

We have also compared the accuracy of a variety of different methods on a benchmark reaction (BHandHLYP, CCSD, HF, M06-2x, CAM-B3LYP and B3LYP) in past work, and found the BH/d computational method provided the best accuracy \citep{Reference598}. We note that the CAM-B3LYP/aug-cc-pVDZ method provided comparable accuracy to BH/d, and would be an interesting method to explore for future network calculations.

Finally, we compare the difference in accuracy for our benchmark method when increasing the size of the basis set from double-$\zeta$ to triple-$\zeta$. We will show that there is no measurable improvement in accuracy for the 12 cases chosen in this work when increasing the basis set size by this level.


The theoretical methods we use to calculate rate coefficients are variational transition state theory (CVT) \citep{Reference534,Reference535,Reference598} and Rice--Ramsperger--Kassel--Marcus/master equation (RRKM/ME) theory \citep{Reference2058}. These theoretical methods are described below.


\subsubsection{One- and Two-Body Reactions}

One- and two-body reaction rate coefficients are calculated using CVT. This method involves varying the reaction coordinate (e.g. a bond distance) along a minimum energy path in order to find the minimum rate coefficient. This is expressed as: \citep{Reference535}

\begin{equation}\label{CVT}
k_{CVT}(T,s) = \min_s \left\lbrace k_{GT}(T,s) \right\rbrace.
\end{equation}
where $k_{GT}(T,s)$ is the generalized transition state theory rate coefficient, $T$ is the temperature, and $s$ is the reaction coordinate (e.g. bond distance).

Neglecting effects due to tunneling, the generalized transition state theory (GT) reaction rate coefficient is given by: \citep{Reference530,Reference535}

\begin{equation}\label{Eyring}
k_{GT}(T,s) = \sigma \frac{k_B T}{h} \frac{Q^{\ddagger}(T,s)}{\prod_{i=1}^{N} Q_i^{n_i}(T)} e^{-E_0(s)/RT}.
\end{equation}
where $\sigma$ is the reaction path multiplicity, {\it k$_B$} is the Boltzmann constant (1.38$\times$10$^{-23}$ J K$^{-1}$), {\it T} is temperature (K), {\it h} is the Planck constant (6.63$\times$10$^{-34}$ J$\cdot$s), {\it Q$^{\ddagger}$} is the partition function of the transition state per unit volume (cm$^{-3}$), with its zero of energy at the saddle point, {\it Q$_i$} is the partition function of species $i$ per unit volume, with its zero of energy at the equilibrium position of species $i$, $n_i$ is the stoichiometric coefficient of species $i$, $N$ is the number of reactant species, {\it E$_0$} is the difference in zero-point energies between the generalized transition state and the reactants (kJ mol$^{-1}$) (0 for barrierless reactions), and {\it R} is the gas constant (8.314$\times$10$^{-3}$ kJ K$^{-1}$ mol$^{-1}$).

To find the location along the minimum energy path where the GT rate coefficient is smallest, we use the maximum Gibbs free energy criterion, which offers a compromise of energetic
and entropic effects \citep{Reference535,Reference538}.

The partition functions per unit volume have four components,

\begin{equation}
Q = \frac{q_t}{V} q_e q_v q_r.
\end{equation}
where V is the volume (cm$^{-3}$) and the $t$, $e$, $v$, and $r$ subscripts stand for translational, electronic, vibrational, and rotational, respectively.

The zero-point energies, Gibbs free energies, and partition functions are calculated for each reaction along its minimum energy path using the Gaussian 09 software package \citep{g09}.

This accurate yet inexpensive method was developed in \citet{Reference598} and typically provides rate coefficients within an order of magnitude of their published experimental values. In this work, in order to improve accuracy by a factor of 2 on average, we modify the location of the zero of energy for the vibrational partition function in Equation~\ref{Eyring}. Now, instead of being at the bottom of the internuclear potential energy well, we place it at the first vibrational level, i.e., the zero-point level. This gives the vibrational partition function the form:

\begin{equation}
q_v = \prod_{n=1}^{N} \frac{1}{1 - e^{-\Theta_n/T}},
\end{equation}
where $N$ is the number of vibrational modes, $\Theta_n$ is the vibrational temperature of the n$^{th}$ mode ($\Theta_n$ = $\frac{\hslash \omega_n}{k_B}$), and T is temperature.


Barrierless reaction rate coefficients do not typically vary by more than a factor of a few for temperatures between 50 and 400 K \citep{Reference587,Reference495,Reference481,Reference577,Reference572}, 
thus temperature dependences are only calculated for the reactions with barriers. This is done by calculating the rate coefficients at 50, 100, 200, 298.15, and 400 K and then fitting the results to the modified Arrhenius expression

\begin{equation}
k(T) = \alpha \left(\frac{T}{300}\right)^{\beta} e^{-\gamma/T},
\end{equation}
where $k(T)$ is the temperature-dependent second-order rate coefficient (cm$^{3}$s$^{-1}$), $\alpha$, $\beta$, and $\gamma$ are fit parameters, and $T$ is temperature (in K).


\subsubsection{Three-Body Reactions}

When two reactants combine to form a single product, a third body is generally required to take away some of the excess vibrational energy from the reaction product, otherwise it will dissociate \citep{Vallance_Book}. The mechanism of these three-body reactions is expressed as

\begin{equation}\label{third-order}
\ce{A + B -> C$_{(\nu)}$}
\end{equation}
\begin{equation}
C_{(\nu)} \xrightarrow{+ M} C.
\end{equation}

In other words, reactants A and B combine to form a vibrationally excited product C$_{(\nu)}$. This product is then collisionally deexcited by species M.

The pressure-dependent rate law for this reaction is

\begin{equation}\label{third_rate_law}
\frac{d[C]}{dt} = k([M]) [A][B].
\end{equation}
where $k([M])$ is the pressure-dependent second-order rate coefficient (cm$^{3}$s$^{-1}$).

In the high atmospheric pressure regime, the collision rate approaches 100\%, and since it cannot exceed this value, the reaction rate becomes pressure independent. Conversely, in the low atmospheric pressure regime, empirical data show that the reaction rate is linear with pressure. For this reason, the pressure dependent rate coefficient is expressed as a function of the Lindemann parameters, i.e., the high-pressure limit ($k_{\infty}$ [cm$^{3}$s$^{-1}$]) and low-pressure limit ($k_0$ [cm$^{6}$s$^{-1}$]) rate coefficients \citep{Reference2064}.


\begin{equation}
k([M]) = \frac{k_0 [M] / k_{\infty}}{1 + k_0 [M] / k_{\infty}} k_{\infty}
\end{equation}


It can be shown by taking the pressure limits of this equation that in the high-pressure limit, $k = k_{\infty}$, and in the low-pressure limit, $k = k_0 [M]$.

We calculate the high-pressure limit rate coefficients in the same way as we calculate two-body rate coefficients, using CVT, however in this case we make use of the ktools code of the Mulitwell Program Suite \citep{multiwell2020,Reference2054,Reference2055}. This method employed in ktools is equivalent to our method of calculating two body rate coefficients; However, we choose to use ktools as it is convenient to building the input files required for calculating low-pressure limit rate coefficients. We find deviations of $<$ 5\% between our manual CVT calculations and those performed by ktools.

To calculate the low-pressure limit rate coefficients, we make use of the Multiwell Master Equation (ME) code, which employs RRKM theory. The ME describes the interaction between collisional energy transfer (with the atmospheric ``bath'' gas) and chemical reaction \citep{Reference2065}. In the case of our three-body reactions, the ME contains the probabilities that our vibrationally excited product will collisionally stabilize for a given atmospheric pressure and temperature. The Multiwell ME code employs Monte Carlo sampling of the ME to build a statistical average of the possible outcomes of the reaction.

The low-pressure limit rate coefficients are then calculated using the output from these stochastic trials: \citep{Reference2054,Reference2061}

\begin{equation}
k_0([M]) = \frac{k_{\infty} f_{prod}}{[M]}
\end{equation}
where $k_{\infty}$ is the high-pressure limit rate coefficient, $f_{prod}$ is the fractional yield of the collisionally deexcited product calculated by the Multiwell ME code, and [M] is the simulated concentration (cm$^{-3}$), which is low enough for $k_0$ to converge.

We used \ce{N2} as a bath gas, as it is the primary constituent of Titan's atmosphere. The energy transfer was treated with a standard exponential-down model with $<\Delta E>_{down}$ = 0.8 T K$^{-1}$ cm$^{-1}$ \citep{Reference2059,Reference2060}. The Lennard-Jones parameters\footnote{The Lennard-Jones potential is used by the ME to model the collision between a molecule and the bath gas.} for \ce{N2} and all the products were taken from the literature \citep{Reference2057,Reference2056,jetsurf2.0} and can be found in Table~\ref{TableLJ}.

In the diffuse, upper regions of atmospheres, vibrationally excited species produced from the combination of two reactants will typically dissociate back into the original reactants. In these cases, the three-body reactions completely describe the chemistry occurring in both diffuse and dense regions of the atmosphere. However in some cases, the favourable vibrational decay products are not the original reactants (e.g. \ce{CN + ^4N -> CN2$_{(\nu)}$* -> N2 + C}). In these cases, we also include the two-body reactions to these favourable decay pathways. 

We would expect these two-body reactions to be less efficient in the lower, denser regions of atmospheres where the vibrationally excited intermediates can be collisionally deexcited. Regardless, we allow these reaction rate coefficients to be independent of pressure. Since radicals are typically not very abundant in the denser regions of the atmosphere, this treatment should not produce significant error.

We verify the most efficient decay pathway of vibrationally excited molecules from previous experimental studies. For more details, we refer the reader to the theoretical case studies in the Appendix materials.



\subsection{Atmospheric Model Parameters}\label{paramSec}

Atmospheric numerical simulations are performed using a 1D chemical kinetic model (ChemKM). Radiative transfer in this code is calculated using the plane-parallel two-stream approximation. Photo-absorption and Rayleigh scattering are also included. ChemKM has been benchmarked with several other chemical kinetic codes\footnote{https://www.issibern.ch/teams/1dchemkinetics/} including Ag{\'u}ndez model \citep{Agundez1,Agundez2}, ARGO \citep{Rimmer_2016}, ATMO \citep{Tremblin_2015,Drummond2016,Tremblin_2017}, Kasting model \citep{Kopparapu_2011,Miguel_2013}, KINETICS \citep{Moses_2011,Moses_2013,Moses_2016}, Venot model \citep{Venot2012,Venot2015}, and VULCAN \citep{Tsai_2017}. Models agree within the numerical precision, when using the same input and setup. For complete details on this atmospheric code, we refer the reader to \citet{Molaverdikhani_2019,Molaverdikhani2020}.

The setup parameters for our fiducial Titan atmospheric model mostly match those in \citet{2012AA...541A..21H}. This includes the atmospheric temperature, pressure, and initial molar mixing ratios of \ce{N2}, \ce{CH4}, \ce{H2}, and \ce{CO}, the eddy diffusion profile, the influx of \ce{H2O} from micrometeorites (5$\times$10$^6$ cm$^{-2}$s$^{-1}$), and 21 of the photochemical reactions and cross-sections. We use the solar mean for the top-of-atmosphere radiation (with solar zenith angle of 50$^{\circ}$) \citep{2004AdSpR..34..256T}. We also initially included the Jean's thermal escape of \ce{H} and \ce{H2}, but found it did not significantly affect our results.

Eddy diffusion describes the turbulent mixing of molecules in an atmosphere. Its form in \citet{2012AA...541A..21H}, which was originally developed by \citet{2008JGRE..11310006H}, is

\begin{equation}
K(z) = \frac{K_o(p_o/p)^{\gamma} K_{\infty}}{K_o(p_o/p)^{\gamma} + K_{\infty}},
\end{equation}
where $K_o$ is the surface eddy coefficient (400 cm$^2$s$^{-1}$), $K_{\infty}$ is the top-of-atmosphere eddy coefficient (3$\times$10$^7$ cm$^2$s$^{-1}$), $p$ is the pressure (Pa), $p_o$ = 1.77$\times$10$^3$ Pa and $\gamma$ = 2.

\begin{figure}[ht!]
\centering
\includegraphics[width=80mm]{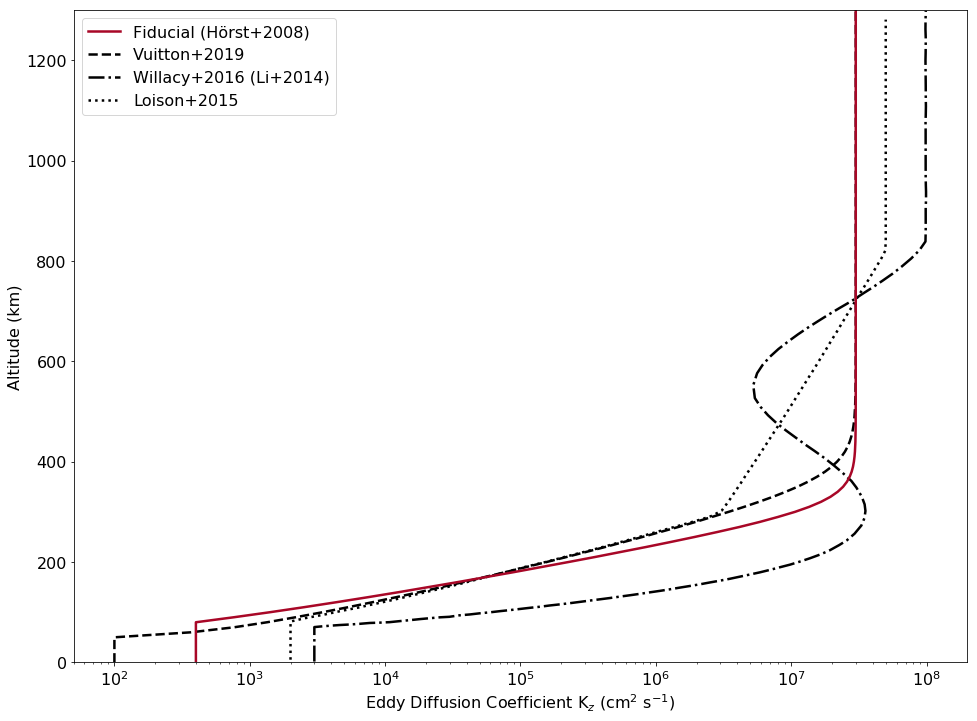}
\caption{Eddy Diffusion profile used in our fiducial model compared with the profiles used in the three most recent Titan models in the literature \citep{2019Icar..324..120V,2016ApJ...829...79W,2015Icar..247..218L}.}
\label{Eddy_Compare}
\end{figure}

In Figure~\ref{Eddy_Compare}, we plot our fiducial eddy diffusion profile along with the profiles of the three most recent Titan models in the literature \citep{2019Icar..324..120V,2016ApJ...829...79W,2015Icar..247..218L}.

Our grid is 100 uniform layers from 0 to 1300 km. 

To handle the galactic cosmic ray (GCR) dissociation of \ce{N2} and \ce{CH4}, we use the most recent GCR models by \citet{2011AA...529A.143G,2009AA...506..955G}. However, we only include the GCR reactions which produce neutral species, as our network does not contain ions. These reactions are, 

\begin{equation*}
\ce{N2 + GCR -> ^2N + ^4N},
\end{equation*}
\begin{equation*}
\ce{N2 + GCR -> ^4N + ^4N},
\end{equation*}
\begin{equation*}
\ce{CH4 + GCR -> CH3 + H},
\end{equation*}
\begin{equation*}
\ce{CH4 + GCR -> ^3CH2 + H2}.
\end{equation*}

We do not include condensation in our models, as we are primarily interested in the gas phase chemistry leading to the production of HCN. In addition, CRAHCN does not include the heavy hydrocarbons which produce the majority of hazes, therefore we do not include haze production/destruction.



\section{Results - Rate Coefficients}\label{results1}

In \citet{Reference598}, we calculated 42 reaction rate coefficients involved with or in competition with the production of HCN in early Earth and Titan atmospheres. In this work, we improve on the original network by including pressure dependence on addition reactions, by modifying the vibrational partition function to improve the accuracy of our calculations (by factor of $\sim$2 on average), and by removing the \ce{2H2CN -> HCN + H2CNH} abstraction reaction, which our recalculations show has a large barrier. We also expand the original network to 104 reactions, which is the result of exploring the entire field of possible reactions for the list of primary species in Table~\ref{Table1}.

In Table~\ref{TableRecalc}, we display the 34 pressure-independent rate coefficients from \citet{Reference598}, recalculated with a modified vibrational partition function described in Section~\ref{methods}. The only pressure-independent reaction not carried over from \citet{Reference598} is \ce{2H2CN -> HCN + H2CNH}. Our recalculations show that this abstraction reaction actually has a large barrier and is therefore too inefficient to consider ($k$(298 K) $\sim$ 10$^{-41}$ cm$^{3}$s$^{-1}$).

20 of these 34 reactions have experimentally measured rate coefficients, and all our calculations land within an order of magnitude of these experimental values. The majority (70\%) land within a factor of 2 of the experimental values. This level of accuracy is consistent with the uncertainties assigned in large-scale experimental data evaluations \citep{Reference451,Reference509}.

We note however that in two cases, our chosen computational method (BH/d) predicted barriers for reactions that did not have barriers. Moreover, in two other cases, this method predicted barrierless reactions for reactions with small experimental barriers. These are limitations of our chosen method, and in these few cases, we artificially remove the barriers from these calculations, or introduce experimental barriers to these calculations, respectively.

As a result of exploring the entire field of possible reactions for the primary species in this work, we calculate the rate coefficients of 36 new two-body reactions, and 32 new three-body reactions (68 total). 33 of these new reactions have no previously known rate coefficient.

In Table~\ref{Table3}, we display the 36 new two-body reactions, along with our calculated rate coefficients at 298 K and any experimentally measured values. Seven of these reactions have experimental values, and the majority (71\%) of our calculated rate coefficients are within a factor of 4 of these values. All our calculated values are within one order of magnitude of experimental values. However, in two cases our chosen computational method predicted no barrier for reactions that have small experimental barriers. As before, we artificially introduce the experimental barriers to these calculations.

In Table~\ref{Table4}, we display the 32 calculated low pressure ($k_0$) and high pressure ($k_{\infty}$) limit rate coefficients at 298 K for the three-body reactions in this work, as well as any experimentally measured values. 16 of the high pressure limit rate coefficients have experimental values, and the majority of cases (69\%) are within a factor of 4 of these experimental values. Again, all calculations are within about an order of magnitude of experimental values.

Our calculated third-order, low pressure limit rate coefficients are within a factor of 2 of experimental values 67\% of the time, and nearly within an order of magnitude in all cases. Only in the case of \ce{CN + ^4N + M -> CN2 + M} is our calculated rate coefficient slightly less accurate, differing from the one experimentally measured value \citep{Reference2046} by a factor of 36. This reaction is not well studied, therefore it is possible that we are not as far off from the exact value as this discrepancy implies. Calculations at the CC/t level of theory only bring this third-order rate coefficient to within a factor of 28 of the experimental value. This reaction turns out not to be important in the story of HCN in Titan's atmosphere.

In Tables~\ref{CRAHCN1} and \ref{CRAHCN2}, we display the 104 pressure- and temperature-dependent rate coefficients for the fiducial chemical network used in this study. Experimentally measured rate coefficients are used when available, which is the case for 42 reactions. Sometimes experimental values are only available for one of either the high-pressure or low-pressure limit rate coefficient, in which case we use a combination of experimental and calculated values. Our calculated values are used in the majority of the network (68\%).



\subsection{Methods Comparison on Dominant Reactions \label{methodcompare}}


In a past computational methods comparison, our calculated rate coefficient for \ce{CH4 + H -> CH3 + H2} at the CCSD/aug-cc-pVDZ level of theory was a factor of 8 smaller than the experimental values \citep{Reference598}. Conversely, our calculated the rate coefficient at the BH/d level of theory was within the experimental range. This, along with speed, were major motivating factors for choosing BH/d for our large-scale theoretical chemical reaction rate study.

In our sensitivity analysis in Section~\ref{sensitivity_analysis}, we find 19 reactions dominate the production and destruction of HCN in Titan's atmosphere. 11 of these 19 reactions have experimental rate coefficients. In the case of \ce{CH3 + H + M -> CH4 + M}, both the high- and low-pressure rate coefficients are experimentally measured.

\begin{table*}[ht!]
\centering
\caption{A comparison of the accuracy of three methods for calculating rate coefficeints for the dominant reactions in this study. The methods are BHandHLYP/aug-cc-pVDZ (BH/d), $\omega$B97XD/aug-cc-pVDZ ($\omega$B/d), and CCSD/aug-cc-pVTZ (CC/t). Only 11 of the 19 dominant reactions have experimentally measured values and can be calculated using our theoretical methods. The error factor is the multiplicative or divisional factor from the nearest experimental or suggested value. For three-body reactions, the displayed rate coefficients are either the high-pressure limit ($k_{\infty}$) or low-pressure limit ($k_0$). The aug-cc-pVDZ basis set is used for all calculations. Second-order rate coefficients have units cm$^{3}$s$^{-1}$. Third-order rate coefficients have units cm$^{6}$s$^{-1}$. \label{CCSD}} 
\begin{tabular}{lcccccccc}
\\
\multicolumn{1}{l}{Reaction equation} & 
\multicolumn{1}{l}{k(298) experiment} & 
\multicolumn{1}{l}{k(298) BH/d} & 
\multicolumn{1}{l}{Error}&
\multicolumn{1}{l}{k(298) $\omega$B/d} & 
\multicolumn{1}{l}{Error}&
\multicolumn{1}{l}{k(298) CC/t} & 
\multicolumn{1}{l}{Error} &
\multicolumn{1}{c}{Winner(s)}
\\ \hline \\[-2mm]
\ce{H2CN + ^4N -> HCN + NH} & 4.4$\times$10$^{-11}$ & $^a$4.7$\times$10$^{-12}$ & 9 & 3.4$\times$10$^{-12}$ & 13 & $^a$5.1$\times$10$^{-12}$ & 9 & $\omega$B\\
\ce{CN + ^4N -> CN2* -> } & 1.0--3.0$\times$10$^{-10}$ & 4.3$\times$10$^{-11}$ & 2 & 4.2$\times$10$^{-11}$ & 2 &  2.6$\times$10$^{-11}$ & 4 & BH, $\omega$B\\
\ce{N2 + C} & & & & & &\\
\ce{CN + CH4 -> HCN + CH3} & 5.6--11$\times$10$^{-13}$ & $^b$7.7$\times$10$^{-13}$ & 1 & $^b$1.3$\times$10$^{-12}$ & 1 &  4.5$\times$10$^{-13}$  & 1 & CC\\
\ce{NH + ^4N -> N2H* -> } & 2.5--2.6$\times$10$^{-11}$ & 5.5$\times$10$^{-11}$ & 2 & 5.6$\times$10$^{-11}$ & 2 &  2.7$\times$10$^{-11}$ & 1 & CC\\
\ce{N2 + H} & & & & & & \\
\ce{^4N + CH3 -> ^3H3CN* -> } & 5.0--7.7$\times$10$^{-11}$ & 6.2$\times$10$^{-11}$ & 1 & 8.7$\times$10$^{-11}$ & 1 & $^c$1.2$\times$10$^{-11}$  & 4 & BH, $\omega$B\\
\ce{H2CN + H} & & & & & & \\
\ce{^2N + CH4 -> H3CNH* -> } & 2.4--4.5$\times$10$^{-12}$ & $^{de}$1.7$\times$10$^{-11}$ & 4 & $^{df}$7.9$\times$10$^{-12}$ & 2 & $^d$2.9$\times$10$^{-11}$ & 6 & tie\\
\ce{^1H2CNH* + H* -> H2CN + H2} & & & & & & \\
\ce{CH4 + ^1CH2 -> C2H6_{(\nu)}* -> } & 0.2--7.3$\times$10$^{-11}$ & 2.1$\times$10$^{-11}$ & 1 & 5.5$\times$10$^{-11}$ & 1 & 1.3$\times$10$^{-11}$ & 1 & tie \\
\ce{CH3 + CH3} & & & & & & \\
\ce{CH4 + CH -> CH4-CH* -> } & 0.02--3$\times$10$^{-10}$ & $^g$1.8$\times$10$^{-10}$ & 1 & 9.2$\times$10$^{-12}$ & 1 & $^g$1.0$\times$10$^{-9}$ & 3 & $\omega$B\\
\ce{C2H5* -> C2H4 + H} & & & & & & \\
\ce{^3CH2 + H -> CH3_{(\nu)}* -> } & 0.8--2.7$\times$10$^{-10}$ & 3.7$\times$10$^{-10}$ & 1 & 7.8$\times$10$^{-10}$ & 3 & 2.6$\times$10$^{-10}$ & 1 & BH, CC\\
\ce{CH + H2} & & & & & & \\
\ce{CH3 + H -> CH4} ($k_{\infty}$) & 1.5--4.7$\times$10$^{-10}$ & 1.4$\times$10$^{-10}$ & 1 & 5.4$\times$10$^{-11}$ & 3 & $^h$2.6$\times$10$^{-10}$ & 1 & BH \\
\ce{CH3 + H + M -> CH4 + M} ($k_0$) & 0.2--5.5$\times$10$^{-28}$ & 2.6$\times$10$^{-28}$ & 1 & 3.3$\times$10$^{-28}$ & 1 & $^h$1.0$\times$10$^{-27}$ & 2 & BH, $\omega$B\\
\ce{H + H + M -> H2 + M } ($k_0$) & 4--250$\times$10$^{-33}$ & 1.7$\times$10$^{-33}$ & 2 & 4.4$\times$10$^{-34}$ & 12 & 1.8$\times$10$^{-33}$ & 2 & BH, CC\\
\hline
\multicolumn{9}{l}{\footnotesize $^a$ We remove the barrier from this calculation as experiments predict this reaction to be barrierless or nearly barrierless \citep{Reference573}.} \\
\multicolumn{9}{l}{\footnotesize See Appendix materials for more details.} \\
\multicolumn{9}{l}{\footnotesize $^b$ We introduce an experimental barrier of 8.3 kJ mol$^{-1}$ \citep{Reference2051} to this calculation as no barrier is found at this level of theory.} \\
\multicolumn{9}{l}{\footnotesize $^c$ We remove the barrier from this calculation as experiments and theory predict this reaction to be barrierless} \\
\multicolumn{9}{l}{\footnotesize \citep{Reference445,Reference442}.} \\
\multicolumn{9}{l}{\footnotesize $^d$ We introduce an experimental barrier of 6.3 kJ mol$^{-1}$ \citep{Reference579} to this calculation as no barrier is found at this level of theory.} \\
\multicolumn{9}{l}{\footnotesize $^e$ Simulations had sporadic convergence beyond a C-N bond distance of 2.76$\AA$. The rate coefficient is calculated with the variational} \\
\multicolumn{9}{l}{\footnotesize transition state at this location, which has the highest $\Delta$G.} \\
\multicolumn{9}{l}{\footnotesize $^f$ Simulations had sporadic convergence beyond a C-N bond distance of 2.60$\AA$. The rate coefficient is calculated with the variational} \\
\multicolumn{9}{l}{\footnotesize transition state at this location, which has the highest $\Delta$G.} \\
\multicolumn{9}{l}{\footnotesize $^g$ We remove the barrier from the rate limiting step of this calculation, i.e. \ce{CH4-CH* -> C2H5*} as experiments predict this reaction} \\
\multicolumn{9}{l}{\footnotesize to be barrierless \citep{Reference561,Reference563,Reference566}.} \\
\multicolumn{9}{l}{\footnotesize $^h$ We remove the barrier from this calculation as experiments and theory predict this reaction to be barrierless \citep{Reference468}.} \\
\end{tabular}
\end{table*}

In Table~\ref{CCSD}, we compare the accuracy of the BH/d, $\omega$B/d, and  CC/t for calculating the rate coefficients of these 11 dominant reactions. Based on general agreement to the experimental values, and a correct diagnosis of the reaction barrier, each method is found to have variable accuracy. In the following paragraphs, we move down through the reactions in Table~\ref{CCSD}, commenting on some of the most important results for selected reactions. 


The results of this methods comparison shows a similar level of consistency in accuracy across all three methods. For the 12 coefficients, BH/d was the most accurate or tied for the most accurate method 8 times, $\omega$B/d was the most accurate or tied for the most accurate method 7 times, and CC/t was the most accurate or tied for the most accurate method 6 times. When comparing only the DFT methods, BH/d was more accurate than $\omega$B/d 3 times, less accurate 2 times, and similarly accurate 7 times.

In some cases, one or more methods would miss a barrier, or find one when one should not be present. In these cases, a method that correctly diagnosed the barrier was considered more accurate than one that incorrectly diagnosed the barrier, regardless of the calculated error factor. Also, all methods that incorrectly diagnosed the barrier were considered equally accurate. CC/t incorrectly diagnosed barriers five times, BH/d incorrectly diagnosed barriers four times, and $\omega$B/d incorrectly diagnosed barriers twice.

In the case of \ce{H2CN + ^4N -> HCN + NH}, the BH/d, $\omega$B/d, and CC/t methods compute barriers of heights E$_0$ $\sim$ 15, 2, and 23 kJ mol$^{-1}$, respectively. The one experimental measurement suggests little or no barrier is present \citep{Reference573} (see Appendix materials for more details). $\omega$B/d computes the smallest barrier; However the rate coefficient calculated using this method is a factor of 13 smaller than the experimental value, removing the barrier brings the calculated rate coefficient to within a factor of 6 of experiment. Given these discrepancies, and the lack of theoretical studies on this reaction, we recommend both a thorough theoretical follow-up study and additional experimental measurements.

Both the BH/d and CC/t methods find the \ce{CH4-CH* -> C2H5*} step of the \ce{CH4 + CH -> C2H4 + H} reaction to have a barrier with a height above the reactants. This differs from our $\omega$B/d calculation, the results of experiment, and other theoretical studies, which suggest this reaction is barrierless \citep{Reference2063,Reference565,Reference561} (see Appendix materials for further details). 


All three methods find no barrier for the \ce{^2N + CH4 -> H3CNH*} reaction step. This step is expected to have a small barrier of E$_0$ = 6.3 kJ mol$^{-1}$ \citep{Reference579}. Similarly, \citet{Reference583} did not find a barrier for this reaction using the CCSD(T)/aug-cc-pVTZ level of theory. 

CC/t misdiagnoses the barrier for the three-body reaction \ce{CH3 + H + M -> CH4 + M}. This reaction is barrierless \citep{Reference468}, however CC/t calculations estimate a barrier of E$_0 \sim$ 62 kJ mol$^{-1}$. Similarly, CC/t misdiagnoses the barrier for \ce{^4N + CH3 -> ^3H3CN*}, which is also barrierless \citep{Reference445,Reference442}. The barrier height for this reaction at the CC/t level of theory is 17 kJ mol$^{-1}$. BH/d and $\omega$B/d correctly calculate no barriers for these two reactions. Conversely, BH/d and $\omega$B/d do not calculate barriers for \ce{CN + CH4 -> HCN + CH3}, which is expected to have a barrier of E$_0$ = 8.3 kJ mol$^{-1}$ \citep{Reference2051}. Our CC/t calculations find a barrier for this reaction of 6.7 kJ mol$^{-1}$.

Lastly, the BH/d and CC/t methods compute rate coefficients for \ce{H + H + M -> H2 + M } that are factors of 2 from the nearest experimental value, whereas the $\omega$B97XD method computes a rate coefficient for this reaction that is a factor of 12 smaller than the nearest experimental value.

Overall, BH/d seems to be a reasonable choice for moving forward with a large scale atmospheric study such as ours, with typical deviations from experiment of a factor of $\leq$ 2. $\omega$B/d would also have been a reasonable choice moving forward, as this method correctly diagnoses barriers more frequently than BH/d and CC/t for this sample size, while maintaining accuracy nearly within an order of magnitude of experimental values. CC/t was the stand alone most accurate method in a two cases, but it was the least accurate method in six cases. Given this, and the much higher computational cost, we do not recommend CC/t for performing rate coefficient calculations for large scale atmospheric studies such as ours. For more comprehensive reaction investigations, we recommend using multiple methods, including CCSD, $\omega$B97XD, and BHandHLYP, to verify the presence or absence of reaction barriers.


\subsection{Basis Set Comparison on Dominant Reactions}

In Table~\ref{basisset}, we compare the accuracy of the aug-cc-pVDZ and aug-cc-pVTZ basis sets, paired with the BHandHLYP method, on the dominant reactions in this study that have experimental rate coefficients. The intent is to see, for our chosen method, if increasing the basis set size from double-$\zeta$ to triple-$\zeta$ leads to an improvement in accuracy with respect to agreement with experimental values.

\begin{table*}[ht!]
\centering
\caption{A comparison of the accuracy of BHandHLYP/aug-cc-pVDZ, and BHandHLYP/aug-cc-pVTZ for calculating rate coefficeints for the difficult dominant reactions in this study. For three-body reactions, the displayed rate coefficients are either the high-pressure limit ($k_{\infty}$) or low-pressure limit ($k_0$). Second-order rate coefficients have units cm$^{3}$s$^{-1}$. Third-order rate coefficients have units cm$^{6}$s$^{-1}$. \label{basisset}} 
\begin{tabular}{lcccccc}
\\
\multicolumn{1}{l}{Reaction equation} & 
\multicolumn{1}{l}{k(298 K) experiment} & 
\multicolumn{1}{l}{k(298 K) BH/d} & 
\multicolumn{1}{l}{Error}&
\multicolumn{1}{l}{k(298 K) BH/t} & 
\multicolumn{1}{l}{Error} &
\multicolumn{1}{l}{\% Difference}
\\ \hline \\[-2mm]
\ce{H2CN + ^4N -> HCN + NH} & 4.4$\times$10$^{-11}$ & $^a$4.7$\times$10$^{-12}$ & 9 & $^a$4.8$\times$10$^{-12}$ & 9 & 2\\
\ce{CN + ^4N -> CN2* -> } & 1.0--3.0$\times$10$^{-10}$ & 4.3$\times$10$^{-11}$ & 2 & 4.1$\times$10$^{-11}$ & 2 & 5\\
\ce{N2 + C} & & & & & & \\
\ce{CN + CH4 -> HCN + CH3} & 5.6--11$\times$10$^{-13}$ & $^b$7.7$\times$10$^{-13}$ & 1 & $^b$5.6$\times$10$^{-13}$ & 1 & 27\\
\ce{NH + ^4N -> N2H* -> } & 2.5--2.6$\times$10$^{-11}$ & 5.5$\times$10$^{-11}$ & 2 & 5.2$\times$10$^{-11}$ & 2 & 5\\
\ce{N2 + H} & & & & & & \\
\ce{^4N + CH3 -> ^3H3CN* -> } & 5.0--7.7$\times$10$^{-11}$ & 6.2$\times$10$^{-11}$ & 1 & 6.2$\times$10$^{-11}$ & 1 & 0\\
\ce{H2CN + H} & & & & & &\\
\ce{^2N + CH4 -> H3CNH* -> } & 2.4--4.5$\times$10$^{-12}$ & $^{cd}$1.7$\times$10$^{-11}$ & 4 & $^{ce}$3.3$\times$10$^{-11}$ & 7& 94\\
\ce{^1H2CNH* + H* -> H2CN + H2} & & & & & & \\
\ce{CH4 + ^1CH2 -> C2H6_{(\nu)}* -> } & 0.2--7.3$\times$10$^{-11}$ & 2.1$\times$10$^{-11}$ & 1 & 2.4$\times$10$^{-11}$ & 1 & 14\\
\ce{CH3 + CH3} & & & & & & \\
\ce{CH4 + CH -> CH4-CH* -> } & 0.02--3$\times$10$^{-10}$ & $^f$1.8$\times$10$^{-10}$ & 1 & $^f$4.1$\times$10$^{-10}$ & 1 & 128\\
\ce{C2H5* -> C2H4 + H} & & & & & & \\
\ce{^3CH2 + H -> CH3_{(\nu)}* -> } & 0.8--2.7$\times$10$^{-10}$ & 3.7$\times$10$^{-10}$ & 1 & 3.7$\times$10$^{-10}$ & 1 & 0\\
\ce{CH + H2} & & & & & &\\
\ce{CH3 + H -> CH4} ($k_{\infty}$) & 1.5--4.7$\times$10$^{-10}$ & 1.4$\times$10$^{-10}$ & 1 & 1.4$\times$10$^{-10}$ & 1 & 0\\
\ce{CH3 + H + M -> CH4 + M} ($k_0$) & 0.2--5.5$\times$10$^{-28}$ & 2.6$\times$10$^{-28}$ & 1 & 2.7$\times$10$^{-28}$ & 1 & 4\\
\ce{H + H + M -> H2 + M } ($k_0$) & 4--250$\times$10$^{-33}$ & 1.7$\times$10$^{-33}$ & 2 & 1.6$\times$10$^{-33}$ & 2 & 6\\
\hline
\multicolumn{7}{l}{\footnotesize $^a$ We remove the barrier from this calculation as experiments predict this reaction to be barrierless or nearly barrierless.} \\
\multicolumn{7}{l}{\footnotesize $^b$ We introduce an experimental barrier of 8.3 kJ mol$^{-1}$ \citep{Reference2051} to this calculation as no barrier is found at this} \\
\multicolumn{7}{l}{\footnotesize level of theory.} \\
\multicolumn{7}{l}{\footnotesize $^c$ We introduce an experimental barrier of 6.3 kJ mol$^{-1}$ \citep{Reference579} to this calculation as no barrier is found at this} \\
\multicolumn{7}{l}{\footnotesize level of theory.} \\
\multicolumn{7}{l}{\footnotesize $^d$ Simulations had sporadic convergence beyond a C-N bond distance of 2.76$\AA$. The rate coefficient is calculated with the variational} \\
\multicolumn{7}{l}{\footnotesize transition state at this location, which has the highest $\Delta$G.} \\
\multicolumn{7}{l}{\footnotesize $^e$ Simulations had sporadic convergence beyond a C-N bond distance of 2.73$\AA$. The rate coefficient is calculated with the variational} \\
\multicolumn{7}{l}{\footnotesize transition state at this location, which has the highest $\Delta$G.} \\
\multicolumn{7}{l}{\footnotesize $^f$ We remove the barrier from the rate limiting step of this calculation, i.e. \ce{CH4-CH* -> C2H5*} as experiments predict this} \\
\multicolumn{7}{l}{\footnotesize reaction to be barrierless \citep{Reference561,Reference563,Reference566}.} \\
\end{tabular}
\end{table*}

What we find, is that rate coefficients calculated at the double-$\zeta$ level are generally very close to the values calculated at the triple-$\zeta$ level. Typical differences are less than 15\%. In two out of twelve cases, the rate coefficients at the double-$\zeta$ level differ from the triple-$\zeta$ values by a factor of $\sim$2. However, in none of these twelve cases is the rate coefficient triple-$\zeta$ level more accurate than the rate coefficient at the double-$\zeta$ level with respect to experimental agreement. For this reason, and considering the added computational cost, we do not upgrade to the triple-$\zeta$ level for our large scale atmospheric study.

\section{Results - HCN in Titan's Atmosphere}\label{results2}

\begin{figure*}[ht!]
\centering
\includegraphics[width=\textwidth]{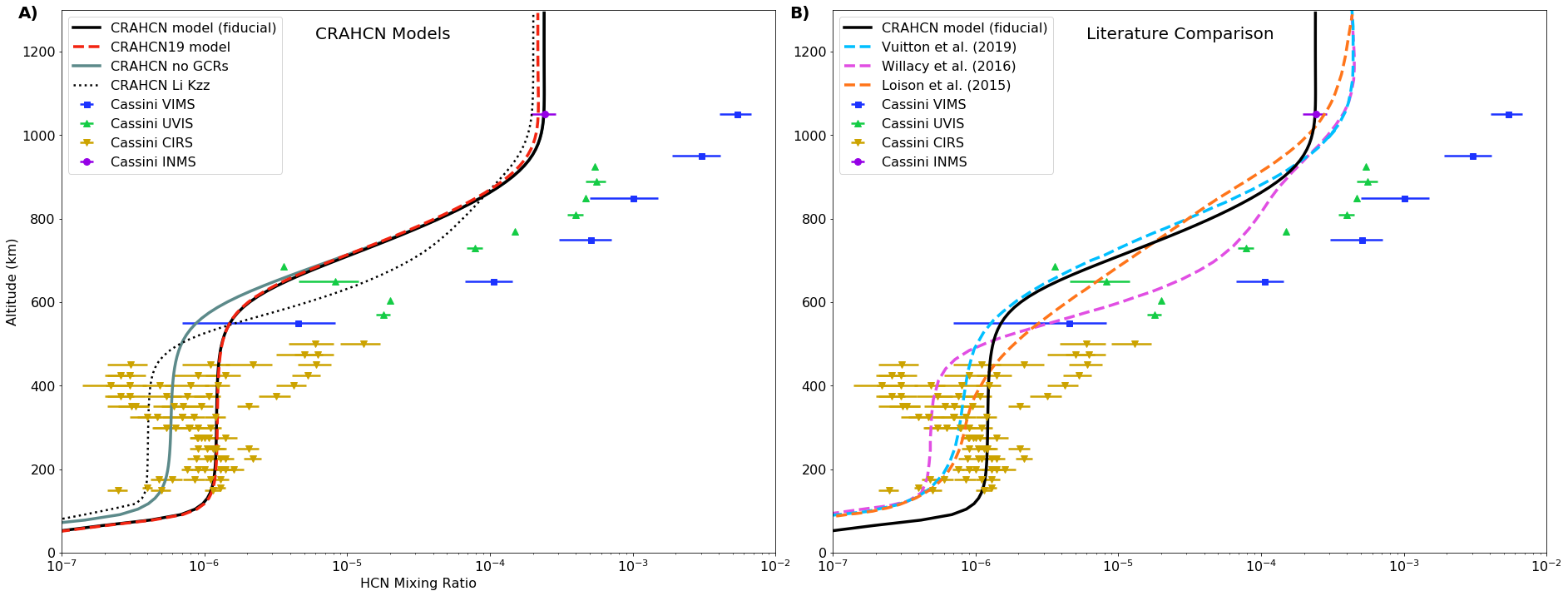}
\caption{Numerical simulations of the HCN molar mixing ratio in Titan's atmosphere compared with observations taken by Cassini. {\bf A)} CRAHCN model: our fiducial model, which uses the CRAHCN network (Tables~\ref{CRAHCN1} and \ref{CRAHCN2}) and fiducial model parameters. CRAHCN19 model: fiducial model parameters, and a chemical network containing only the dominant 19 reactions. CRAHCN no GCRs: fiducial model parameters and the CRAHCN chemical network, but all GCR reactions are turned off. CRAHCN Li Kzz: same as the CRAHCN no GCRs model, except we use the eddy diffusion profile from \citet{2014PSS..104...48L} instead of the fiducial one from \citet{2008JGRE..11310006H}. The data points represent observations taken by the Cassini spacecraft. The spread in the Cassini CIRS data is due to measurements taken at various latitudes \citep{2010Icar..205..559V}. {\bf B)} CRAHCN Model comparison with the three most recent Titan atmospheric chemistry models in the literature \citep{2019Icar..324..120V,2016ApJ...829...79W,2015Icar..247..218L}. For the literature model parameters, see Table~\ref{modelParams}.}
\label{HCNTitanAtmosphere}
\end{figure*}

In Figure~\ref{HCNTitanAtmosphere}A, we display our 4 modeled atmospheric HCN profiles for Titan, as well as the HCN observations made in Titan's atmosphere by the Cassini spacecraft. Each model is discussed in detail in the subsections below. In Figure~\ref{HCNTitanAtmosphere}B, we compare our fiducial HCN profile to those of the three most recent Titan models in the literature \citep{2019Icar..324..120V,2016ApJ...829...79W,2015Icar..247..218L}. 

\subsection{CRAHCN model (fiducial)}

For our fiducial model, we use the CRAHCN network and the fiducial model parameters as described in Section~\ref{paramSec}. In Figure~\ref{HCNTitanAtmosphere}A, we see that the HCN profile from our fiducial model agrees very well with the HCN observations in Titan's lower atmosphere, landing right in the middle of the Cassini CIRS measurements. Our fiducial profile also nails the single Cassini INMS data point at 1050 km, which is the only {\it in situ} HCN measurement of Titan's atmosphere. The trade-off in agreeing so well with the INMS measurement, is that we do not agree as well with the VIMS limb measurements, or the UVIS stellar occultation measurements. This, as can be seen in Figure~\ref{HCNTitanAtmosphere}B, is standard for current state-of-the-art Titan models.

\subsection{CRAHCN19 model}

Our sensitivity analyses of the CRAHCN network (discussed in Section~\ref{sensitivity_analysis} below) revealed that 19 reactions are predominantly involved in the production and destruction of HCN in Titan's atmosphere. For the CRAHCN19 model, we use the fiducial model parameters, and a network containing only the 19 dominant reactions out of the 104 total reactions in CRAHCN. In Figure~\ref{HCNTitanAtmosphere}A, we see the HCN profile from the CRAHCN19 model almost perfectly aligns with the profile from the CRAHCN model. Maximum deviations between these model curves in the upper atmosphere are $\sim$10\%. The total HCN produced in the CRAHCN19 model is only 6\% more than the HCN produced in the fiducial model. This is mainly due to slight deviations in the lower atmosphere between models. This result suggests that the CRAHCN19 network contains nearly all that is necessary to simulate the production of HCN in Titan-like atmospheres.

\subsection{CRAHCN no GCRs model}

The aim of the CRAHCN no GCRs model is to examine the sensitivity of HCN production in the lower atmosphere to GCR flux. Thus, this model is similar to our fiducial model, except that all GCR reactions are removed. In Figure~\ref{HCNTitanAtmosphere}A, we see the HCN profile from the CRAHCN no GCRs model overlaps with the fiducial model in the upper atmosphere, where no GCR reactions occur. The HCN profile in the mid-lower atmosphere is only reduced by a factor of $\sim$2 compared to the fiducial model. Overall, the CRAHCN no GCRs model produces about one-third as much HCN in Titan's atmosphere as the fiducial model. This result suggests that GCRs are not critical for the production of HCN in Titan's atmosphere, however they boost total HCN production by about a factor of $\sim$3.

\subsection{CRAHCN Li Kzz model}

To investigate the effects of eddy diffusivity in the distribution of atmospheric HCN, we present a model in which we modify the eddy diffusion profile to match that of \citet{2014PSS..104...48L} (see Figure~\ref{Eddy_Compare}). This profile differs from the \citet{2008JGRE..11310006H} profile used in all other models in that the profile inverts at 350 km, creating a low eddy diffusion zone near 550 km. This is also the profile used in the \citet{2016ApJ...829...79W} Titan model. In Figure~\ref{HCNTitanAtmosphere}A, we see that using this eddy diffusion profile reduces the HCN abundance in the lower atmosphere by a factor of $\sim$3, and increases the HCN abundance in the mid atmosphere by a factor of $\sim$4 with respect to our fiducial model. Overall, the HCN profile from this model does not agree with the Cassini CIRS data as well as our fiducial model, as the former misses the range of CIRS measurements from 200--300 km by about a factor of 3.

\subsection{Comparison to Other Recent Titan Models}

In Figure~\ref{HCNTitanAtmosphere}B, we plot HCN profiles from the three most recent Titan models \citep{2019Icar..324..120V,2016ApJ...829...79W,2015Icar..247..218L} to compare with our fiducial HCN profile. It is important to emphasize that the three models from the literature focused on reproducing the observed profiles of many chemical species, only one of which was HCN. The differences in parameters and chemistry between these models is summarized in Table~\ref{modelParams} and Figure~\ref{Eddy_Compare}.

\begin{table*}[ht!]
\centering
\caption{Summary of the major differences in model parameters and chemistry between our fiducial model and the three most recent Titan models in the literature \citep{2019Icar..324..120V,2016ApJ...829...79W,2015Icar..247..218L}. There are multiple Titan models in \citet{2019Icar..324..120V} and \citet{2016ApJ...829...79W}: we choose the models that best agree with the Cassini HCN measurements. \label{modelParams}} 
\begin{tabular}{lcccc}
\\
\multicolumn{1}{l}{Model} & 
\multicolumn{1}{l}{Reaction Network} & 
\multicolumn{1}{l}{Photolytic processes} &
\multicolumn{1}{l}{Eddy diffusion} & 
\multicolumn{1}{l}{GCR processes} 
\\ \hline \\[-2mm]
Fiducial (this work) & 104 & 21 & Figure~\ref{Eddy_Compare} & N$_2$ and CH$_4$ \\
\citet{2019Icar..324..120V} (K$_o$=100) & $>$ 3000 reactions  & 116 & '' & N$_2$ and CH$_4$ \\
\citet{2016ApJ...829...79W} (Model A/B) & not listed & not listed & '' & none \\
\citet{2015Icar..247..218L} & 969  & 171 & '' & N$_2$ only \\
\hline
\end{tabular}
\end{table*}

Differences between all four model curves are within a factor of $\sim$3 in the lower atmosphere, a factor of $\sim$8 in the mid atmosphere, and a factor of $\sim$2 in the upper atmosphere. Given the differences in eddy diffusion profiles, condensation/sedimentation, photochemistry, GCR chemistry, and reaction networks, a complete explanation on the variations between these curves is not possible, however, we note a few things below.

The HCN profile from \citet{2016ApJ...829...79W} varies the most from our fiducial model. However, comparing the HCN model profile from \citet{2016ApJ...829...79W} to our CRAHCN Li Kzz model in Figure~\ref{HCNTitanAtmosphere}A, we can see that the curves have a very similar form in the lower and mid atmosphere. Therefore, we suspect the major differences between our fiducial HCN profile and the HCN profile in \citet{2016ApJ...829...79W} to be due to differences in eddy diffusion.

The HCN model from \citet{2019Icar..324..120V} varies from our fiducial HCN model by $\lesssim$2. \citet{2019Icar..324..120V} parameterized eddy diffusion in the same way we do, however we use a slightly higher surface eddy coefficient (K$_o$ = 400 cm$^2$s$^{-1}$ versus 100 cm$^2$s$^{-1}$, see Figure~\ref{Eddy_Compare}). \citet{2019Icar..324..120V} analyzed how changes to their surface eddy coefficient affected their HCN profile, and found that shifting K$_o$ = 100 cm$^2$s$^{-1}$ to 1000 cm$^2$s$^{-1}$ decreased their HCN content in the lower atmosphere. This suggests that the major differences between our fiducial HCN profile and the HCN profile in \citet{2019Icar..324..120V} are due to differences in chemical networks and photochemistry, rather than eddy diffusion.

Differences between our fiducial HCN profile and the HCN profile in \citet{2015Icar..247..218L} also vary by $\lesssim$2. However, due to lack of data, we cannot comment on a major source of the discrepancies.

Other differences between our model and those in the literature include treatments for condensation/sedimentation and haze formation. We do not include condensation/sedimentation in our fiducial model, as we are mainly interested in the gas phase chemistry leading to the production of HCN. In addition, CRAHCN does not include the heavy hydrocarbons that produce the majority of hazes. \citet{2016ApJ...829...79W} find that condensation/sedimentation only affects the HCN profile below $\sim$100 km, which is below any Cassini measurement. \citet{2016ApJ...829...79W} also included the permanent removal of HCN via haze production in one of their models (Model C), which resulted in a reduction of HCN below $\sim$500 km of approximately a factor of 4 compared to their models without hazes (Models A and B). Their haze model, however, does not agree with the Cassini CIRS HCN data as well as their models without haze production.

Overall, our fiducial HCN model is in general agreement with the most recent Titan models in the literature. 

\subsection{Sensitivity Analyses}\label{sensitivity_analysis}

Not every reaction in an atmospheric chemical network contributes significantly to the production and destruction of a given species. To discover which reactions in CRAHCN contributes to the fiducial HCN profile we perform two types of sensitivity analyses on our fiducial model. 

\subsubsection{Sensitivity Analysis I}

The first sensitivity analysis involves running 104 additional numerical simulations of Titan's atmosphere. In each simulation, one of the 104 reactions in CRAHCN is removed, and the resultant HCN profile compared with the fiducial HCN profile. We also perform this sensitivity analysis on the CRAHCN/no GCRs model.

\begin{figure*}[ht!]
\centering
\includegraphics[width=0.7\textwidth]{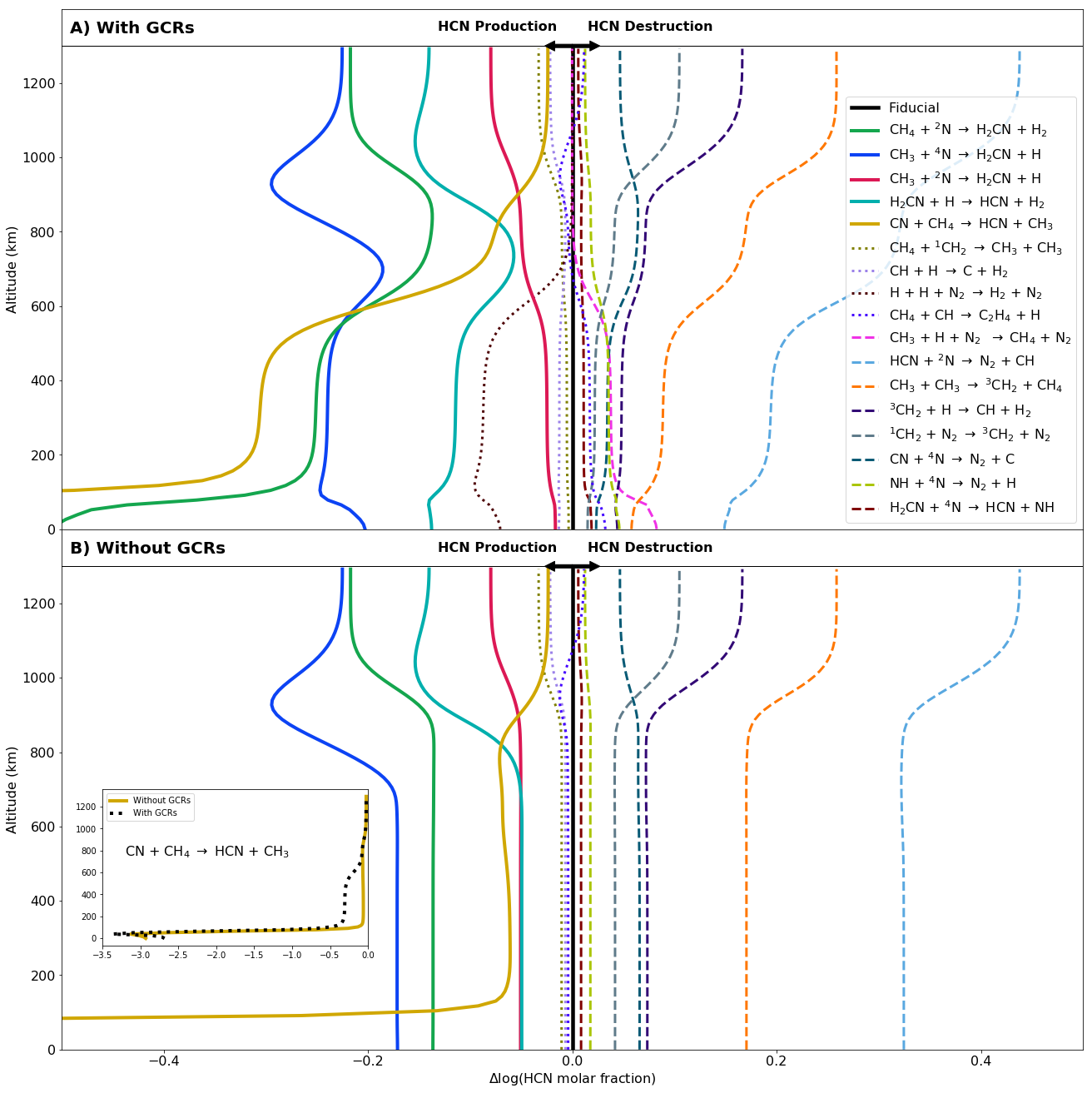}
\caption{Sensitivity analysis I revealing 17 of the dominant 19 reactions in Titan's atmosphere. Each reaction curve shows the difference in HCN molar mixing ratio when excluding that reaction from {\bf A)} the fiducial model, and {\bf B)} the CRAHCN/no GCRs model. All of the other reactions in CRAHCN did not greatly affect the HCN profile upon their exclusion; for both models.}
\label{Sensitivity}
\end{figure*}

This first sensitivity analysis revealed 17 of the 19 dominant reactions. In Figure~\ref{Sensitivity}A we display the changes to the HCN profiles that occur when each of these 17 reactions are removed. The removal of all other reactions did not significantly effect the fiducial HCN profile.

\subsubsection{Sensitivity Analysis II}

The second sensitivity analysis involves running a much larger number of simulations. Starting with all 104 reactions, as was done in sensitivity analysis I, each reaction is excluded in a simulation to see how it effects the HCN profile. The reaction whose removal affects the HCN profile the least is then removed, and the process is repeated with 103 reactions. The least important reaction is removed at each stage, until the exclusion of any of the remaining reactions leads to a $\gtrsim$10\% deviation from the fiducial HCN profile.

\begin{figure*}[ht!]
\centering
\includegraphics[width=0.7\textwidth]{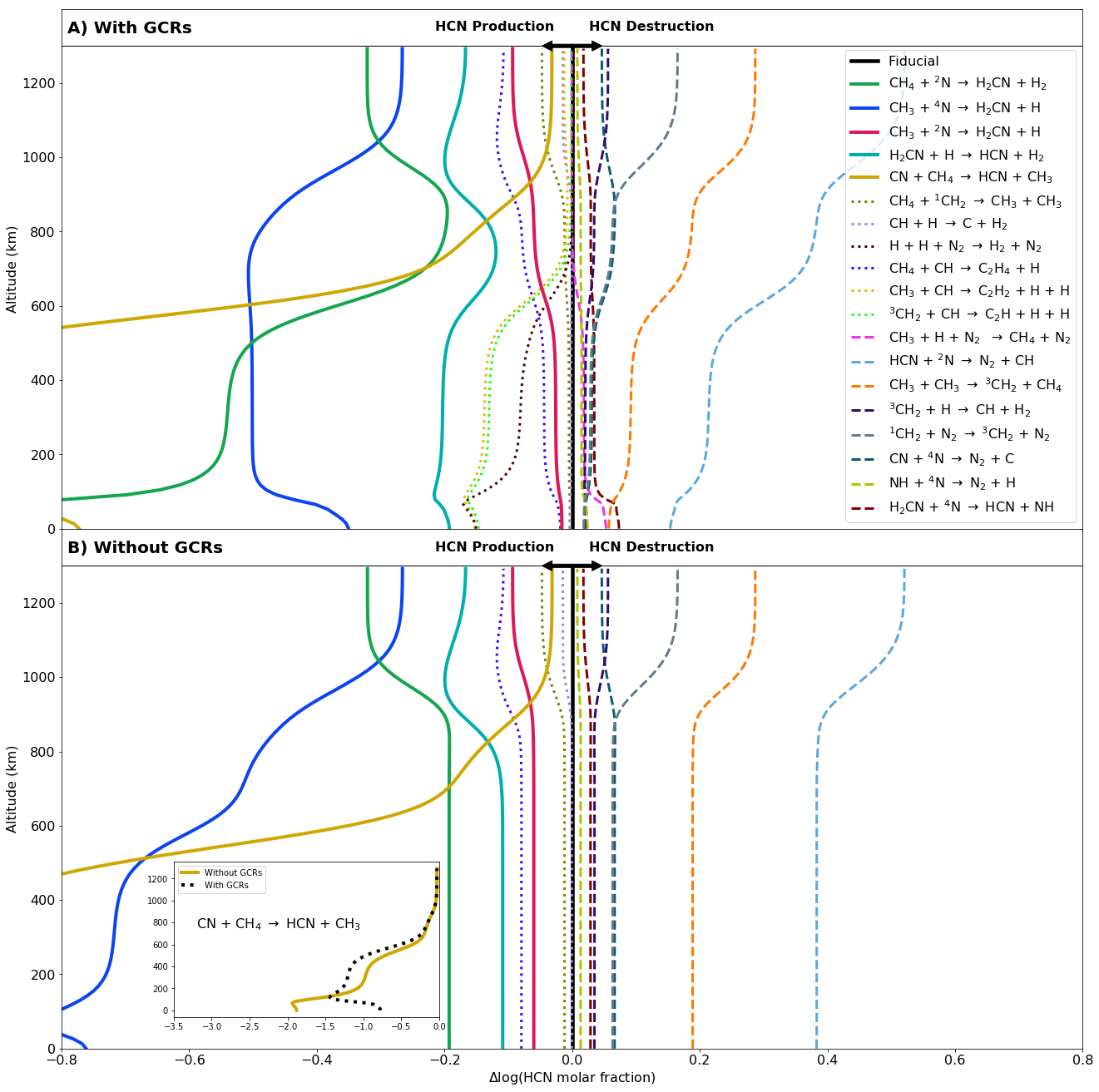}
\caption{Sensitivity analysis II revealing all 19 dominant reactions in Titan's atmosphere. Each reaction curve shows the difference in HCN molar mixing ratio when excluding that reaction from {\bf A)} the CRAHCN19 model, and {\bf B)} the CRAHCN19/no GCRs model. All of the other reactions in CRAHCN did not greatly affect the HCN profile upon their exclusion; for both models.}
\label{Sensitivity2}
\end{figure*}

The second sensitivity analysis revealed 2 additional dominant reactions, \ce{CH3 + CH -> C2H2 + H + H} and \ce{^3CH2 + CH -> C2H + H + H}, bringing the total to 19 dominant reactions. In Figure~\ref{Sensitivity2}A we display the changes to the HCN profiles that occur when each of these 19 reactions are removed from a network containing only these 19 reactions (CRAHCN19).

\subsubsection{19 Dominant Reactions}

The 19 dominant reactions are listed in Table~\ref{19reac} below. Five reactions dominate the production of HCN, four are critical for increasing the feedstock of precursor molecules that react to produce HCN, one reaction dominates the destruction of HCN, seven reactions reduce the key precursor molecules that produce HCN, one reaction attenuates the precursor sinks by reducing H abundance, and one reaction acts as both a precursor source and sink.

\begin{table*}[ht!]
\centering
\caption{The 19 reactions responsible for the production and destruction of HCN in Titan's atmosphere, labelled with their dominant role. For simplicity, reaction intermediates are not listed here. See Tables~\ref{CRAHCN1} and \ref{CRAHCN2} for full details of reaction intermediates. Reactions are considered ``well studied'' if they have more than one experimental measurement or theoretical study at room temperature \label{19reac}} 
\begin{tabular}{lclc}
\\
\multicolumn{1}{l}{Role} & 
\multicolumn{1}{l}{No.} & 
\multicolumn{1}{l}{Reaction equation} &
\multicolumn{1}{l}{Well studied?}\\ \hline \\[-2mm]
HCN sources & 54. & \ce{CN + CH4 -> HCN + CH3} & Yes\\
& 73. & \ce{^2N + CH4 -> H2CN + H2}  & Yes\\
& 69. & \ce{^4N + CH3 -> H2CN + H}  & Yes\\
& 74. & \ce{^2N + CH3 -> H2CN + H}  & No\\
& 43. & \ce{H2CN + H -> HCN + H2}  & No\\
 & & & \\
Precursor & 82. & \ce{CH4 + ^1CH2 -> CH3 + CH3} & Yes\\
sources& 89. & \ce{CH3 + CH -> C2H2 + H + H} & No \\
& 95. &  \ce{^3CH2 + CH -> C2H + H + H} & No \\
& 104. & \ce{CH + H -> C + H2} & No \\
& & & \\
HCN sink & 46. & \ce{HCN + ^2N -> N2 + CH} & No \\
 & & & \\
Precursor & 81. & \ce{CH3 + CH3 -> ^3CH2 + CH4} & No \\
sinks & 34. & \ce{^1CH2 + N2 -> ^3CH2 + N2} & Yes \\
& 52. & \ce{CN + ^4N -> N2 + C} & Yes \\
& 96. & \ce{^3CH2 + H -> CH + H2} & Yes \\
& 37. & \ce{H2CN + ^4N -> HCN + NH} & No \\
& 22. & \ce{CH3 + H + N2 -> CH4 + N2} & Yes\\
& 61. & \ce{NH + ^4N -> N2 + H} & Yes\\
& & & \\
Precursor sink & 32. & \ce{H + H + N2 -> H2 + N2} & Yes \\
attenuation & & & \\
& & & \\
Precursor & & & \\
source/sink  & 83. & \ce{CH4 + CH -> C2H4 + H} & Yes\\
\hline
\end{tabular}
\end{table*}

The biggest impact to the fiducial HCN profile is the removal of \ce{CN + CH4 -> HCN + CH3}. This is the key reaction that recycles CN---primarily from HCN photodissociation---back into HCN. This reaction accounts for $\sim$36--46\%\footnote{Percent contributions for the four main HCN channels are calculated by dividing the difference in the total HCN abundance (integrated over all altitudes) when removing that reaction, by the summed up total differences in HCN abundances when removing each of the four main reactions. Calculations differ when using the fiducial and CRAHCN19 models, therefore we express the values as a range.} of the total HCN in Titan's atmosphere, and is dominant primarily because of Titan's high atmospheric \ce{CH4} abundance.

The next most important channel is \ce{^2N + CH4 -> H2CN + H2}, followed by \ce{H2CN + H -> HCN + H2}. Again, due primarily to the high atmospheric \ce{CH4} concentrations, this multi-step reaction is responsible for $\sim$32--38\% of the total HCN in Titan's atmosphere. 

The next leading reaction for HCN production is \ce{^4N + CH3 -> H2CN + H -> HCN + H2}, which accounts for $\sim$20--25\% of the total HCN in Titan's atmosphere.

Finally, the final dominant reaction for HCN production in Titan's atmosphere, is \ce{^2N + CH3 -> H2CN + H -> HCN + H2}. This reaction was discovered by \citet{Reference598}, and had no known rate coefficient prior to that work. It accounts for $\sim$2\% of the total HCN in Titan's atmosphere.

It is worth noting that the leading reaction, \ce{CN + CH4 -> HCN + CH3}, only produces HCN if CN is present. CN primarily comes from the photodestruction of HCN, therefore this reaction is not responsible for starting HCN synthesis in Titan's atmosphere, but rather, maintaining it. Given that we ignore this maintenance reaction, the other three leading channels produce approximately 59\%, 37\%, and 4\% of the total \emph{initial} HCN in Titan's atmosphere, respectively.

There are four reactions that play an important role in processing radical species to produce the precursors for HCN production reactions. The main one in the upper atmosphere is \ce{CH4 + ^1CH2 -> CH3 + CH3}, which provides \ce{CH3} for reactions 69 and 74. Although this reaction removes a \ce{CH4} molecule, which is also a reactant for \ce{HCN} production via reactions 54 and 73, the rate coefficients for these \ce{CH4}-based reactions are 2--5 orders of magnitude smaller than those for reactions 69 and 74, and thus producing more \ce{CH3} leads to more efficient HCN production. The other upper atmospheric precursor source is \ce{CH + H -> C + H2} followed by the photolysis of \ce{H2} to form two \ce{H} atoms to be used by reaction 43.

The other two processing reactions, which were only revealed by the second sensitivity analysis, produce the \ce{H} atoms necessary for \ce{H2CN + H -> HCN + H2}. These reactions dominate in the lower atmosphere, where UV light does not reach and thus \ce{H2} photodissociation does not occur.


The dominant sink for HCN is \ce{HCN + ^2N -> N2 + CH} and accounts for nearly 100\% of total HCN removal. Although the photodissociation reaction \ce{HCN + $h\nu$ -> CN + H} destroys HCN efficiently, the removal of this reaction does not significantly affect the total HCN abundance in Titan's atmosphere. This is because CN efficiently reacts with \ce{CH4} to recycle back into HCN.

Several reactions reduce HCN production by acting as sinks to important HCN precursors, i.e., \ce{CH3}, \ce{CN}, \ce{CH4}, \ce{^1CH2}, \ce{^3CH2}, \ce{^4N}, and \ce{H}. One of these precursor sink reactions (no. 37) ironically produces \ce{HCN}. However, since this reaction also produces \ce{NH}, the dominant effect is the removal of \ce{^4N} from the atmosphere via \ce{NH + ^4N -> N2 + H}. One reaction, \ce{CH4 + CH -> C2H4 + H}, seems to act as a precursor sink in the lower atmosphere, and a precursor source in the mid-upper atmosphere. However, this changes when switching from the fiducial to the CRAHCN19 models, and therefore the true role of this reaction is uncertain.

Finally, one reaction is key to attenuating the effect of a precursor sink in the lower atmosphere. \ce{H + H + N2 -> H2 + N2} reduces the \ce{H}-atom abundance to attenuate the effects of \ce{CH3 + H + N2 -> CH4 + N2}.

\subsection{The Case of No GCRs}

In Figures~\ref{Sensitivity}B and \ref{Sensitivity2}B, we display the changes to the HCN profiles that occur when each the 19 dominant reactions are removed from the CRAHCN/no GCRs and CRAHCN19/no GCRs models, respectively. As a reminder, the 19 dominant reactions are the reactions which upon their removal, have the greatest effect the HCN profile. Interestingly, only 15 reactions, upon their removal, affected the no GCR HCN profiles. The removal of \ce{CH3 + CH -> C2H2 + H + H}, \ce{^3CH2 + CH -> C2H + H + H}, \ce{CH3 + H + N2 -> CH4 + N2} and \ce{H + H + N2 -> H2 + N2} did not significantly effect the CRAHCN/no GCRs or CRAHCN19/noGCRs HCN profiles.

These four particular reactions require high abundances of \ce{CH3}, \ce{^3CH2}, and \ce{H} produced by the GCR destruction of \ce{CH4} in the lower atmosphere in order to become important for HCN production and destruction. It can be seen in Figures~\ref{Sensitivity}A and \ref{Sensitivity2}A that these four reactions have no affect on the HCN profiles in the upper atmosphere, where GCR reactions do not occur.

In the upper atmospheres, the HCN difference profiles in Figures~\ref{Sensitivity}B and \ref{Sensitivity2}B look nearly identical to the like-colored profiles in Figure~\ref{Sensitivity}A. Differences between any two like-colored curves are more drastic in the lower atmospheres, where GCR reactions occur. Removing GCR reactions changes the feedstock of methane and nitrogen radicals, and therefore adjusts the relative importance of each of the dominant reactions that use these radicals.

The dominant pathways to HCN formation in the CRAHCN/no GCRs models are the same as those in the fiducial model, however their percent contributions differ by up to 20\% from the fiducial model values. \ce{CN + CH4 -> HCN + CH3} in the no GCRs case contributes $\sim$42--52\% to the total HCN in Titan's atmosphere. The other three channels, i.e. \ce{^2N + CH4 + H -> H2CN + H2 + H -> HCN + 2H2}, \ce{^4N + CH3 -> H2CN + H -> HCN + H2}, and \ce{^2N + CH3 -> H2CN + H -> HCN + H2}, contribute $\sim$16--18\%, $\sim$22--36\%, and $\sim$6--8\%, respectively.

Overall, these results suggest that GCR reactions do not significantly control which reactions dominate at producing and destroying HCN in Titan's atmosphere, but they do affect the relative amount that they contribute to the overall HCN abundance.

\section{Discussion}\label{discuss}

\subsection{How HCN is Produced in Titan's Atmosphere}

Out of the 104 chemical reactions in CRAHCN, we find only 19 reactions significantly contribute to the production and destruction of HCN in Titan's atmosphere. Most of these reactions are direct sources and sinks for HCN and sources and sinks for the precursors to HCN (e.g. \ce{CH3, ^4N}). In the only other case, a reaction has the role of attenuating the effects of a precursor sink.

\begin{figure*}[ht!]
\centering
\includegraphics[width=0.7\textwidth]{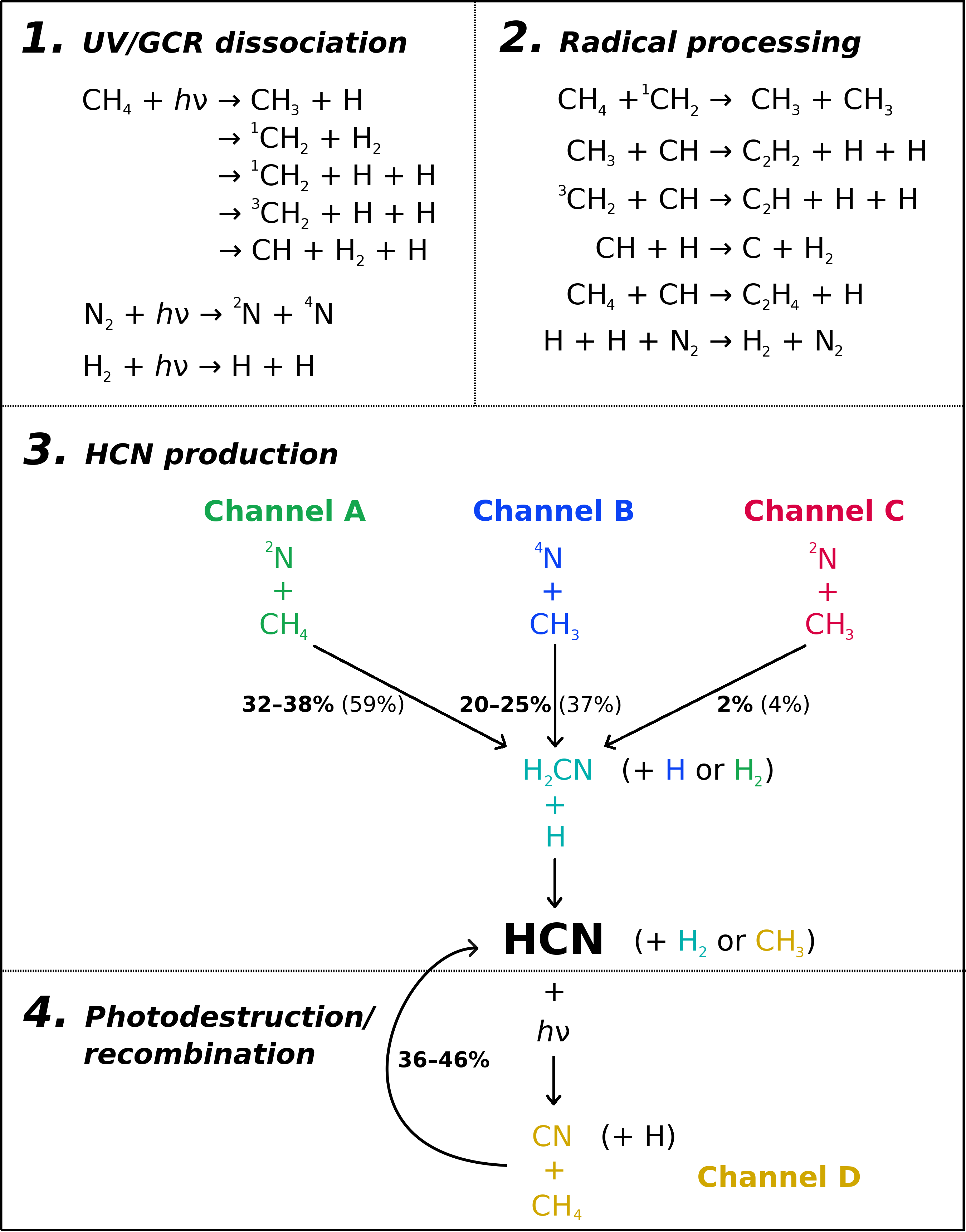}
\caption{Summary of how HCN is produced in Titan's atmosphere. Stage 1: Destruction of methane, nitrogen and hydrogen by ultraviolet light in the upper atmosphere and galactic cosmic rays (GCRs) in the lower atmosphere. Stage 2: Increasing the abundance of HCN precursors and attenuating the effects of HCN precursor sinks. Stage 3: Production of HCN from methane and nitrogen fragments. Stage 4: HCN photodestruction to produce \ce{CN}, and recycling of \ce{CN} back into HCN via reaction with \ce{CH4}. Bold percentages include all HCN reaction routes, including CN recombination after photodestruction. Percentages in parentheses represent \emph{initial} HCN production and do not include CN recombination.}
\label{HCNSchematic}
\end{figure*}

In Figure~\ref{HCNSchematic}, we describe the step-by-step process of HCN production in Titan's atmosphere. First, UV radiation in the upper atmosphere, and GCRs in the lower atmosphere, break apart \ce{CH4}, \ce{N2}, and \ce{H2} into reactive high-energy radical species. Second, these radicals get processed via chemical reactions to form HCN precursors (e.g. \ce{CH3} and \ce{H}). These processing reactions differ in the upper and lower atmosphere. For example, \ce{CH4 + ^1CH2 -> CH3 + CH3} is only a key processing reaction in the upper atmosphere where the reactant \ce{^1CH2} is produced from the UV dissociation of \ce{CH4}. Conversely, \ce{CH3 + CH -> C2H2 + H + H} is only important in the lower atmosphere where a key alternate \ce{H}-atom source (UV dissociation of \ce{H2}) does not occur.

Somewhat unintuitively, too much \ce{H} in the lower atmosphere can lead to less HCN, as it is a reactant in the precursor sink reaction \ce{CH3 + H + N2 -> CH4 + N2}. For this reason, the precusor sink attenuation reaction \ce{H + H + N2 -> H2 + N2} is also important for increasing HCN production in the lower atmosphere.

Next, HCN production occurs through 3 main channels, with the total \emph{initial} percent contributions labeled in parentheses

\begin{multline*}
\textbf{Channel A:} \\
\ce{^2N + CH4 + H -> H2CN + H2 + H -> HCN + 2H2}(59\%),
\end{multline*}
\begin{multline*}
\textbf{Channel B:}  \\
\ce{^4N + CH3 -> H2CN + H -> HCN + H2} \quad (37\%),
\end{multline*}
\begin{multline*}
\textbf{Channel C:}  \\
\ce{^2N + CH3 -> H2CN + H -> HCN + H2} \quad (4\%).
\end{multline*}

In the upper atmosphere, where partial pressures are low, UV radiation is the main dissociating agent; it generates reactants for Channels A--C. Channel B dominates the HCN production in this region, due to its comparatively high rate coefficient. Here, UV radiation is also responsible for breaking apart HCN into \ce{CN + H}. Eddy diffusion mixes species including CN from the upper atmosphere into the lower atmosphere, where Channels A and D mainly take over HCN production. In the lower atmosphere, high partial pressures screen out UV radiation and increase the probability for GCR collisions, therefore GCRs become the main dissociating agent here. In this region, concentrations of \ce{CH4} are high enough that the relative reaction rate of Channel A surpasses that of Channel B. These high concentrations of \ce{CH4} also drastically increase the reaction rate of Channel D, which recycles CN back into HCN. This recycling process is the overall dominant channel to HCN, accounting for 36--46\% of the total HCN in Titan's atmosphere. Channel C remains the fourth most important HCN source in both areas of the atmosphere, as although the rate coefficient of \ce{^2N} reacting with \ce{CH3} is higher than that of \ce{^2N} reacting with \ce{CH4}, there is a much higher concentration of \ce{CH4} compared with \ce{CH3} in all areas of the atmosphere.

\subsection{Using CRAHCN}

Due to the exceptional alignment of the HCN profiles from the CRAHCN19 and CRAHCN models, we suggest the CRAHCN19 network provides a lean, accurate, fast, and intuitively clear code to calculate the HCN abundance in Titan-like atmospheres. Without data from higher pressure and temperature planetary atmospheres, we cannot be certain that these same 19 reactions would suffice for other planetary environments of Titan-like composition. For this reason, we advocate using the full (104 reaction) CRAHCN network to simulate the production of HCN in \ce{N2}-, \ce{CH4}-, and \ce{H2}-dominated atmospheres. 

We emphasize that this is a reduced network to accurately model HCN chemistry, rather than an extended network to cover the chemistry of a large range of species. For this reason, CRAHCN should only be used to simulate the production HCN.

\section{Conclusions}\label{conclusions}

In this paper, we calculate the production of HCN in Titan's atmosphere using a novel quantum chemistry and atmospheric modeling strategy. This strategy has two components: 1) we use quantum chemistry simulations to scan the entire field of possible reactions for a list of primary species relevant to \ce{N2}-, \ce{CH4}-, and \ce{H2}-dominated atmospheres. We then calculate the rate coefficients for the uncovered reactions and construct a consistent reduced atmospheric hybrid chemical network (CRAHCN). This network contains experimental rate coefficients when available (32\% of cases), but is predominantly composed of our calculated values using a consistent computational and theoretical method. 2) We pair CRAHCN with a chemical kinetic code called ChemKM to model the atmosphere of Titan. HCN has been observed at a range of altitudes in Titan's atmosphere by the Cassini spacecraft, making it an excellent testbed for validating chemical networks for HCN production in atmospheres.

We list the major conclusions of this work in bullet form below.

\begin{itemize}
\item CRAHCN contains 104 reactions, 33 of which are newly discovered in this work.
\item Our calculated rate coefficients are accurate to within about an order of magnitude of experimental values, which is consistent with the uncertainties assigned in large-scale experimental data evaluations.
\item In comparison with other widely used computational quantum methods, BHandHLYP/aug-cc-pVDZ is found to provide a reasonable balance of speed and acceptable accuracy for our large scale atmospheric study. Increasing the basis set to aug-cc-pVTZ did not improve the accuracy of calculations with respect to agreement with experimental values.
\item The HCN profile from our fiducial model of Titan's atmosphere agrees very well with the Cassini observations, and is well in line with the three most recent Titan models in the literature.
\item Only 19 reactions are responsible for the production and destruction of HCN in Titan's atmosphere. These reactions are sources and sinks of HCN, sources and sinks of the precursors to HCN, and a presursor sink attenuation reaction.
\item There are 4 main channels to HCN production:
\begin{itemize}
\item \ce{CN + CH4 -> HCN + CH3} ($\sim$36--46\%),
\item \ce{^2N + CH4 + H -> H2CN + H2 + H -> HCN + 2H2} ($\sim$32--38\%),
\item \ce{^4N + CH3 -> H2CN + H -> HCN + H2} ($\sim$20--25\%),
\item \ce{^2N + CH3 -> H2CN + H -> HCN + H2} ($\sim$2\%).
\end{itemize}
\item The first and second reactions dominate in the lower atmosphere, whereas the second, third, and fourth reactions dominate in the upper atmosphere. In the upper atmosphere, where partial pressures are low, relatively high rate coefficients tend to dictate the dominant reactions. In the lower atmosphere, the high \ce{CH4} partial pressure increases the reaction rates for the first two reactions, which is the reason for their dominance here.
\item The fourth dominant source of HCN in Titan's atmosphere is a reaction first discovered in our recent work \citep{Reference598}.
\item \ce{HCN + ^2N -> N2 + CH} is the main sink for HCN. Conversely, \ce{HCN + $h\nu$ -> CN + H} is not an effective HCN sink, because it produces a \ce{CN} molecule that reacts with \ce{CH4} to form back into \ce{HCN}.
\item GCRs triple the total production of HCN in Titan's atmosphere, however they do not affect which reactions dominate HCN production and destruction.
\end{itemize}

Our work suggests that chemical networks of hundreds or thousands of reactions are not necessary to accurately simulate the production of HCN in \ce{N2}-, \ce{CH4}-, and \ce{H2}-dominated atmospheres. Instead, using our novel strategy of exploring the entire field of possible reactions for a short list of primary atmospheric species has proven to be valuable at uncovering the dominant chemical pathways to producing HCN in Titan's atmosphere.

In upcoming work, we will use this strategy to expand CRAHCN to explore the production of HCN in the early Earth atmosphere, which, along with \ce{N2}, \ce{CH4}, and \ce{H2}, is expected to have contained oxygen-based primary species such as \ce{CO2} and \ce{H2O}, as well as their dissociation fragments.

\section*{Acknowledgments}

We thank the two anonymous referees, whose detailed reports led to improvements to this work. We gratefully acknowledge Profs. Paul Ayers (McMaster) and Sarah H{\"o}rst (Johns Hopkins University), whose advice and comments were valuable to this research. B.K.D.P. is supported by an
NSERC Canada Graduate Scholarships-Doctoral (CGS-D). R.E.P. is supported by an NSERC Discovery Grant. T.H. acknowledges support from the European Research Council under the Horizon 2020 Framework Program via the ERC Advanced Grant Origins 83 24 28. We are grateful to Compute Canada for the computer time
allocated for this research. This research is part of a collaboration between the Origins Institute and the Heidelberg Initiative for the Origins of Life.

\beginsupplement

\bibliography{Bibliography_Titan}
\bibliographystyle{aasjournal}

\section*{Appendix Materials}

\subsection*{Rate Coefficient Calculations}

In Table~\ref{TableRecalc}, we display the recalculated two-body reaction rate coefficients from \citet{Reference598}, with the modified vibrational partition function described in Section~\ref{methods}.

\setlength\LTcapwidth{\textwidth}
\begin{longtable*}{lccccc}
\caption{Re-calculated two-body reaction rate coefficients at 298 K from \citet{Reference598} with new vibrational partition function model. For these calculations, the vibrational partition functions are adjusted so that the zero of energy is at the first vibrational level, i.e., the zero-point level. Calculations are performed at the BHandHLYP/aug-cc-pVDZ level of theory. The presence or absence of an energy barrier in the rate-limiting step of the reaction is specified. The error factor is the multiplicative or divisional factor from the nearest experimental or suggested value. \label{TableRecalc}} \\
Reaction equation & Forw./Rev. & Barrier? & k(298) calculated & k(298) experimental & Error factor\\ \hline \\[-2mm]
\ce{H2CN + ^4N -> HCN + NH} & F & Y & $^a$4.7$\times$10$^{-12}$ & 4.4$\times$10$^{-11}$ & 9 \\
\ce{H2CN + H -> HCN + H2} & F & Y & 2.2$\times$10$^{-11}$ & 8.3$\times$10$^{-11}$ & 4 \\
\ce{H2CN -> HCN + H} & F & Y &  2.2$\times$10$^{-15}$ & & \\
\ce{NH + ^4N -> N2H* -> N2 + H} & F & N & 5.5$\times$10$^{-11}$ & 2.5--2.6$\times$10$^{-11}$ & 2 \\
\ce{NH + ^2N -> N2H* -> N2 + H} & F & N & 8.8$\times$10$^{-11}$ & & \\
\ce{NH + H -> H2 + ^4N} & F & Y & 7.1$\times$10$^{-12}$ & 3.2$\times$10$^{-12}$ & 2 \\
\ce{^4N + CH3 -> ^3H3CN* -> H2CN + H} & F & N & 6.2$\times$10$^{-11}$ & 5--7.7$\times$10$^{-11}$ & 1 \\
\ce{^4N + ^3CH2 -> H2CN$_{(\nu)}$* -> HCN + H} & F & N & 1.0$\times$10$^{-10}$ & & \\
\ce{^4N + ^1CH2 -> ^4H2CN -> ^3HCN* + H* ->} & F & N & 1.9$\times$10$^{-10}$ & & \\
\ce{CN + H + H} & & & & & \\
\ce{^4N + CH -> ^3HCN* -> CN + H} & F & N & 1.5$\times$10$^{-10}$ & 0.2--1.6$\times$10$^{-10}$ & 1\\
\ce{^2N + CH4 -> H3CNH* -> ^1H2CNH* + H* -> } & F & N & $^{bX}$1.7$\times$10$^{-11}$ & 2.4--4.5$\times$10$^{-12}$ & 4 \\
\ce{H2CN + H2} & & & & & \\
\ce{^2N + CH3 -> ^{3,1}H3CN* -> ^{3,1}H2CNH*} & F & N & 2.3$\times$10$^{-10}$ & & \\
\ce{H2CN + H} & & & & & \\
\ce{^2N + ^3CH2 -> H2CN$_{(\nu)}$* -> HCN + H} & F & N & 2.1$\times$10$^{-10}$ & & \\
\ce{^2N + ^3CH2 -> ^4H2CN* -> ^3HCN* + H* -> } & F & N & 3.5$\times$10$^{-10}$ & & \\
\ce{CN + H + H} & & & & & \\
\ce{^2N + ^1CH2 -> H2CN$_{(\nu)}$* -> HCN + H} & F & N & 2.4$\times$10$^{-10}$ & & \\
\ce{^2N + CH -> ^3HCN* -> CN + H} & F & N & 3.5$\times$10$^{-10}$ & & \\
\ce{^2N + H2 -> NH2$_{(\nu)}$* -> NH + H} & F & N & $^c$3.3$\times$10$^{-11}$ & 1.7--5.0$\times$10$^{-12}$ & 7 \\
\ce{CH4 + ^3CH2 -> CH3 + CH3} & F & Y & 5.7$\times$10$^{-18}$ & $<$3-500000$\times$10$^{-19}$ & 1 \\
\ce{CH4 + ^3CH2 <- CH3 + CH3} & R & N & 5.3$\times$10$^{-11}$ & & \\
\ce{CH4 + ^1CH2 -> C2H6$_{(\nu)}$* -> CH3 + CH3} & F & N & 2.1$\times$10$^{-11}$ & 0.2--7.3$\times$10$^{-11}$  & 1\\
\ce{CH4 + CH -> CH4-CH* -> C2H5* -> } & F & N & $^d$1.8$\times$10$^{-10}$ & 0.02--3$\times$10$^{-10}$ & 1 \\
\ce{C2H4 + H} & & & & & \\
\ce{CH4 + H -> CH3 + H2} & F & Y & 8.9$\times$10$^{-19}$ & 8.2--350$\times$10$^{-19}$ & 1 \\
\ce{CH4 + H <- CH3 + H2} & R & Y & 3.3$\times$10$^{-20}$ & 9.6--13$\times$10$^{-21}$ & 3 \\
\ce{CH3 + ^3CH2 -> C2H5* -> C2H4 + H} & F & N & 5.9$\times$10$^{-11}$ & 5--21$\times$10$^{-11}$ & 1 \\
\ce{CH3 + ^1CH2 -> C2H5* -> C2H4 + H} & F & N & 1.3$\times$10$^{-10}$ & 3.0$\times$10$^{-11}$ & 2 \\
\ce{CH3 + H -> ^3CH2 + H2} & F & Y & 8.1$\times$10$^{-21}$ & & \\
\ce{CH3 + H <- ^3CH2 + H2} & R & Y & 5.9$\times$10$^{-16}$ & $<$5--50$\times$10$^{-15}$ & 1 \\
\ce{^3CH2 + ^3CH2 -> C2H4$_{(\nu)}$* -> C2H2 + H2} & F & N & 3.6$\times$10$^{-11}$ & 5.3$\times$10$^{-11}$ & 1\\
\ce{^3CH2 + ^1CH2 -> ^3C2H4* -> C2H3* + H -> } & F & N & 2.1$\times$10$^{-10}$ & 3.0$\times$10$^{-11}$ & 7\\
\ce{C2H2 + H + H} & & & & & \\
\ce{^1CH2 + ^1CH2 -> C2H4$_{(\nu)}$* -> C2H2 + H2} & F & N & 7.1$\times$10$^{-11}$ & 5.0$\times$10$^{-11}$ & 1\\
\ce{^1CH2 + H2 -> CH4$_{(\nu)}$* -> CH3 + H} & F & N & 1.0$\times$10$^{-10}$ & 0.07--1.3$\times$10$^{-10}$ & 1 \\
\ce{CH + CH -> ^3C2H2* -> C2H + H} & F & N & 1.4$\times$10$^{-10}$ &  &  \\
\ce{CH + H2 -> CH3$_{(\nu)}$* -> ^3CH2 + H} & F & N & 5.4$\times$10$^{-11}$ & 0.1--16$\times$10$^{-11}$ & 1 \\
\ce{CH + H -> ^3CH2$_{(\nu)}$* -> C + H2} & F & N & 6.9$\times$10$^{-10}$ & & \\
\hline
\multicolumn{6}{l}{\footnotesize $^a$ We remove the barrier from this calculation as experiments predict this reaction to be barrierless or nearly barrierless \citep{Reference573}. } \\
\multicolumn{6}{l}{\footnotesize See case study 1 for more details.} \\
\multicolumn{6}{l}{\footnotesize $^b$ The existence of a barrier for this reaction is dependent on computational method. We introduce an experimental barrier of 6.3 kJ mol$^{-1}$} \\
\multicolumn{6}{l}{\footnotesize \citep{Reference579} to this calculation as at the BHandHLYP/aug-cc-pVDZ level of theory, we find this reaction to be barrierless.} \\
\multicolumn{6}{l}{\footnotesize $^X$ Simulations had sporadic convergence beyond a C-N bond distance of 2.76$\AA$. The rate coefficient is calculated with the variational} \\
\multicolumn{6}{l}{\footnotesize transition state at this location, which has the highest $\Delta$G.} \\
\multicolumn{6}{l}{\footnotesize $^c$ The existence of a barrier for this reaction is dependent on computational method. We introduce an experimental barrier of 7.3 kJ mol$^{-1}$} \\
\multicolumn{6}{l}{\footnotesize \citep{Reference552} to this calculation as at the BHandHLYP/aug-cc-pVDZ level of theory, we find this reaction to be barrierless.} \\
\multicolumn{6}{l}{\footnotesize $^d$ We remove the barrier from the rate limiting step of this calculation, i.e. \ce{CH4-CH* -> C2H5*} as experiments predict this reaction} \\
\multicolumn{6}{l}{\footnotesize to be barrierless \citep{Reference561,Reference563,Reference566}.} \\
\end{longtable*}

In Table~\ref{Table3}, we list the new two-body reaction rate coefficients calculated in this work at 298 K, along with any experimental values.

\setlength\LTcapwidth{\textwidth}
\begin{longtable*}{lccccc}
\caption{New two-body reaction rate coefficients calculated in this work at 298 K. Calculations are performed at the BHandHLYP/aug-cc-pVDZ level of theory. Reactions with rate coefficients slower than k = 10$^{-21}$ cm$^{3}$s$^{-1}$ are not included in this network. The presence or absence of an energy barrier in the rate-limiting step of the reaction is specified. The error factor is the multiplicative or divisional factor from the nearest experimental or suggested value. \label{Table3}} \\
Reaction equation & Forw./Rev. & Barrier? & k(298) calculated & k(298) experimental & Error factor\\ \hline \\[-2mm]
\ce{H2CN + NH -> HCN + NH2} & F & Y & 1.4$\times$10$^{-13}$ & & \\
\ce{H2CN + ^4N -> ^3H2CNN* -> N2 + ^3CH2} & F & N & 4.3$\times$10$^{-12}$ & & \\
\ce{H2CN + ^2N -> ^3H2CNN* -> N2 + ^3CH2} & F & N & 3.2$\times$10$^{-11}$ & & \\
\ce{H2CN + ^2N -> ^1H2CNN* -> N2 + ^1CH2} & F & N & 6.4$\times$10$^{-12}$ & & \\
\ce{H2CN + CH -> ^1H2CNCH* ->} & F & Y & 1.4$\times$10$^{-11}$ & & \\ 
\ce{ H2CNC_a* + H* -> H2CNC_b* + H* -> } & & & & & \\
\ce{CH2CN + H} & & & & & \\ 
\ce{H2CN + CH -> ^3H2CNCH* -> HCN + ^3CH2} & F & N & 2.2$\times$10$^{-11}$ & &\\
\ce{H2CN + CH -> ^1H2CNCH* -> HCN + ^1CH2} & F & Y & 3.3$\times$10$^{-11}$ & & \\
\ce{HCN + CN -> HNCCN* -> NCCN + H} & F & N & 3.7$\times$10$^{-12}$ & 0.2--4.1$\times$10$^{-13}$ & 9\\
\ce{HCN + ^2N -> N2 + CH} & F & N & 6.8$\times$10$^{-11}$ & & \\
\ce{HCN + ^1CH2 -> CH2HCN* -> } & F & N & 3.7$\times$10$^{-13}$ & & \\
\ce{CH2NC_a* + H* -> CH2NC_b* + H* -> } & & & & & \\
\ce{CH2CN + H} & & & & & \\
\ce{HCN + H -> HCNH* -> HNC + H} & F & Y & 2.7$\times$10$^{-20}$ & & \\
\ce{HCN + H <- HCNH* <- HNC + H} & R & Y & 1.0$\times$10$^{-11}$ & & \\
\ce{CN + NH -> HCN + ^4N} & F & Y & 1.9$\times$10$^{-14}$ & & \\
\ce{CN + NH -> HNNC* -> CHN2_a* -> } & F & Y & 1.9$\times$10$^{-12}$ & & \\
\ce{CHN2_b* -> CH + N2} & & & & & \\
\ce{CN + ^4N -> CN2$_{(\nu)}$* -> N2 + C} & F & N & 4.3$\times$10$^{-11}$ & 1.0--3.0$\times$10$^{-10}$ & 2 \\
\ce{CN + ^2N -> CN2$_{(\nu)}$* -> N2 + C} & F & N & 1.6$\times$10$^{-10}$ & & \\
\ce{CN + CH4 -> HCN + CH3} & F & Y & $^a$7.7$\times$10$^{-13}$ & 5.6--11$\times$10$^{-13}$ & 1 \\
\ce{CN + CH3 -> HCN + ^3CH2} & F & N & 6.7$\times$10$^{-12}$ & & \\
\ce{CN + CH -> HCN + C} & F & N & 1.4$\times$10$^{-11}$ & & \\
\ce{CN + CH -> HNC + C} & F & N & $^b$5.4$\times$10$^{-13}$ & & \\
\ce{CN + H2 -> HCN + H} & F & Y & $^c$1.3$\times$10$^{-15}$ & 1.2--4.9$\times$10$^{-14}$ & 9 \\
\ce{NH + NH -> N2H2* -> N2H* + H* ->} & F & N & 4.1$\times$10$^{-12}$ & 3.5$\times$10$^{-12}$ & 1\\
\ce{N2 + H + H} & & & & & \\
\ce{NH + NH -> NH2 + ^4N} & F & Y & 6.8$\times$10$^{-18}$ &  &\\
\ce{NH + CH3 -> H3CNH* -> } & F & N & 2.0$\times$10$^{-11}$ & & \\
\ce{^1H2CNH* + H* -> H2CN + H2} & & & & & \\
\ce{NH + ^3CH2 -> ^1H2CNH* -> H2CN + H} & F & N & 1.9$\times$10$^{-11}$ & & \\
\ce{NH + ^1CH2 -> ^3H2CNH* -> H2CN + H} & F & N & 4.6$\times$10$^{-11}$ & & \\
\ce{NH + CH -> HNCH* -> HCN + H} & F & N & $^d$4.5$\times$10$^{-11}$ & & \\
\ce{NH + CH -> HNCH* -> HNC + H} & F & Y & $^d$3.9$\times$10$^{-15}$ & & \\
\ce{CH3 + CH -> ^1CH3CH* -> C2H3* + H* ->} & F & N & 5.5$\times$10$^{-11}$ & & \\ 
\ce{C2H4} & & & & & \\
\ce{CH3 + CH -> ^3CH3CH* -> C2H3* + H* ->} & F & N & 2.4$\times$10$^{-10}$ & & \\
\ce{C2H2 + H + H} & & & & & \\
\ce{^3CH2 + CH -> C2H3* -> C2H2 + H} & F & N & 6.3$\times$10$^{-11}$ & & \\
\ce{^3CH2 + CH -> ^4C2H3* -> ^3C2H2* + H* ->} & F & N & 1.1$\times$10$^{-10}$ & & \\
\ce{C2H + H + H} & & & & &\\
\ce{^3CH2 + H -> CH3$_{(\nu)}$* -> CH + H2} & F & N & 3.7$\times$10$^{-10}$ & 0.8--2.7$\times$10$^{-10}$ & 1\\
\ce{^1CH2 + CH -> C2H3* -> C2H2 + H} & F & N & 1.6$\times$10$^{-10}$ & & \\
\ce{^1CH2 + H -> CH3$_{(\nu)}$* -> CH + H2} & F & N & 1.8$\times$10$^{-10}$ & 5.0$\times$10$^{-11}$ & 4 \\
\ce{^1CH2 + H -> CH3$_{(\nu)}$* -> ^3CH2 + H} & F & N & 4.6$\times$10$^{-11}$ & & \\
\hline
\multicolumn{6}{l}{\footnotesize $^a$ The existence of a barrier for this reaction is dependent on computational method. We introduce an experimental barrier of 8.3 kJ mol$^{-1}$ } \\
\multicolumn{6}{l}{\footnotesize \citep{Reference2051} to this calculation as at the BHandHLYP/aug-cc-pVDZ level of theory, we find this reaction to be barrierless.} \\
\multicolumn{6}{l}{\footnotesize $^b$ Simulations had sporadic convergence beyond a H-N bond distance of 1.45$\AA$. The rate coefficient is calculated with the variational} \\
\multicolumn{6}{l}{\footnotesize transition state at this location, which has the highest $\Delta$G.} \\
\multicolumn{6}{l}{\footnotesize $^c$ The existence of a barrier for this reaction is dependent on computational method. We introduce an experimental barrier of 13.4 kJ mol$^{-1}$} \\
\multicolumn{6}{l}{\footnotesize \citep{Reference2053} to this calculation as at the BHandHLYP/aug-cc-pVDZ level of theory, we find this reaction to be barrierless.} \\
\multicolumn{6}{l}{\footnotesize $^d$ Simulations did not converge beyond a N-C bond distance of 3.33$\AA$. The rate coefficient is calculated with the variational transition} \\
\multicolumn{6}{l}{\footnotesize state at this location, which has the highest $\Delta$G.} \\
\end{longtable*}

In Table~\ref{Table4}, we display the reduced Lindemann parameters for the three-body reactions calculated in this work, along with any experimental values.

\begin{longtable*}{lcccccc}
\caption{Lindemann coefficients for the three-body reactions calculated in this work at 298 K. Calculations are performed at the BHandHLYP/aug-cc-pVDZ level of theory, and are valid within the 50--400 K temperature range. $k_{\infty}$ is the second-order rate coefficient in the high pressure limit with units cm$^{3}$s$^{-1}$. $k_0$ is the third-order rate coefficient in the low pressure limit with units cm$^{6}$s$^{-1}$. These values fit into the pressure-dependent rate coefficient equation $k$ = $\frac{k_0 [M] / k_{\infty}}{1 + k_0 [M] / k_{\infty}} k_{\infty}$.  Reactions with rate coefficients slower than k$_{\infty}$ = 10$^{-13}$ cm$^{3}$s$^{-1}$ are not included in this network. The error factor is the multiplicative or divisional factor from the nearest experimental or suggested value. \label{Table4}} \\
Reaction equation & k$_\infty$(298) calc. & k$_\infty$(298) exp. & Error$_\infty$ & k$_0$(298) calc. & k$_0$(298) exp. & Error$_0$ \\ \hline \\[-2mm]
\ce{HCN + CH + M -> HCNCH_a* + M* -> } & 7.1$\times$10$^{-11}$ & 2.7$\times$10$^{-10}$ &  4 & 3.3$\times$10$^{-29}$ & & \\
\ce{HCNCH_b* + M* -> CH2NC* + M* ->} & & & & & & \\
\ce{CH2CN + M} & & & & & & \\
\ce{CN + CN + M -> NCCN + M} & 1.6$\times$10$^{-12}$ & 1.0--9.4$\times$10$^{-12}$ & 1 & 5.3$\times$10$^{-29}$ & 4.7--4900$\times$10$^{-32}$ & 1 \\
\ce{CN + CN + M -> CNCN + M} & 6.3$\times$10$^{-12}$ &  &  & 2.2$\times$10$^{-30}$ & &\\
\ce{CN + NH + M -> HNCN + M} & 2.3$\times$10$^{-11}$ & & & 5.9$\times$10$^{-29}$ & &\\
\ce{CN + ^4N + M -> CN2 + M} & 4.3$\times$10$^{-11}$ & 1.0--3.0$\times$10$^{-10}$ & 2 & 1.0$\times$10$^{-30}$ & 2.8$\times$10$^{-32}$ & 36 \\
\ce{CN + ^2N + M -> CN2 + M} & 1.6$\times$10$^{-10}$ & & & 1.1$\times$10$^{-29}$ & &\\
\ce{CN + CH3 + M -> CH3CN + M} & 1.3$\times$10$^{-11}$ & & & 3.8$\times$10$^{-26}$ & & \\
\ce{CN + ^3CH2 + M -> CH2CN + M} & 2.9$\times$10$^{-11}$ & & & 8.5$\times$10$^{-27}$ & & \\
\ce{CN + ^1CH2 + M -> CH2CN + M} & 6.3$\times$10$^{-11}$ & & & 3.5$\times$10$^{-26}$ & & \\
\ce{CN + CH + M -> ^3HCCN + M} & 1.1$\times$10$^{-11}$ & & & 6.5$\times$10$^{-28}$ & &\\
\ce{CN + CH + M -> ^1HCCN + M -> } & 2.9$\times$10$^{-11}$ & & & 1.2$\times$10$^{-28}$ & &\\
\ce{^1HC2N + M} & & & & & & \\
\ce{CN + H + M -> HCN + M} & 4.5$\times$10$^{-11}$ & 1.7$\times$10$^{-10}$ & 4 & 1.7$\times$10$^{-30}$ & 1.3--1.6$\times$10$^{-30}$ & 1\\
\ce{^4N + ^3CH2 + M -> H2CN + M} & 1.0$\times$10$^{-10}$ & & & 1.2$\times$10$^{-28}$ & & \\
\ce{^4N + ^4N + M -> N2 + M} & 1.4$\times$10$^{-11}$ & & & 4.2$\times$10$^{-34}$ & 2.2--130$\times$10$^{-33}$ & 5\\
\ce{^4N + H + M -> NH + M} & 4.3$\times$10$^{-10}$ & & & 3.2$\times$10$^{-33}$ & 1.4--64$\times$10$^{-33}$ & 1\\
\ce{^2N + ^3CH2 + M -> H2CN + M} & 2.1$\times$10$^{-10}$ & & & 6.2$\times$10$^{-28}$ & &\\
\ce{^2N + ^1CH2 + M -> H2CN + M} & 2.4$\times$10$^{-10}$ & & & 1.5$\times$10$^{-27}$ & & \\
\ce{^2N + ^2N + M -> N2 + M} & 5.7$\times$10$^{-11}$ & & & 1.7$\times$10$^{-33}$ & &\\
\ce{^2N + H + M -> NH + M} & 7.7$\times$10$^{-10}$ & & & 7.0$\times$10$^{-32}$ & &\\
\ce{CH4 + ^1CH2 + M -> C2H6 + M} & 2.1$\times$10$^{-11}$ & 0.2--7.3$\times$10$^{-11}$ & 1 & 7.2$\times$10$^{-24}$ & &\\
\ce{CH3 + CH3 + M -> C2H6 + M} & 5.9$\times$10$^{-12}$ & 3.5--6.5$\times$10$^{-11}$ & 6 & 1.7$\times$10$^{-26}$ & 1.7--3.3$\times$10$^{-26}$ & 1 \\
\ce{CH3 + H + M -> CH4 + M} & 1.4$\times$10$^{-10}$ & 1.5--4.7$\times$10$^{-10}$ & 1 & 2.6$\times$10$^{-28}$ & 1.5--55$\times$10$^{-29}$ & 1\\
\ce{^3CH2 + ^3CH2 + M -> C2H4 + M} & 3.5$\times$10$^{-11}$ & 5.3$\times$10$^{-11}$ & 2 & 2.6$\times$10$^{-26}$ & &\\
\ce{^3CH2 + H + M -> CH3 + M} & 4.6$\times$10$^{-10}$ & 8.3--27$\times$10$^{-11}$ & 2 & 4.8$\times$10$^{-29}$ & &\\
\ce{^1CH2 + ^1CH2 + M -> C2H4 + M} & 7.1$\times$10$^{-11}$ & 5.0$\times$10$^{-11}$ & 1 & 5.4$\times$10$^{-25}$ & &\\
\ce{^1CH2 + H2 + M -> CH4 + M} & 1.0$\times$10$^{-10}$ & 7--130$\times$10$^{-12}$ & 1 & 1.4$\times$10$^{-27}$ & &\\
\ce{^1CH2 + H + M -> CH3 + M} & 2.3$\times$10$^{-10}$ & 5.0$\times$10$^{-11}$ & 5 & 3.7$\times$10$^{-28}$ & &\\
\ce{CH + CH + M -> C2H2 + M} & 1.2$\times$10$^{-11}$ & 1.7--2.0$\times$10$^{-10}$ & 14 & 3.5$\times$10$^{-28}$ & &\\
\ce{CH + H2 + M -> CH3 + M} & 2.7$\times$10$^{-10}$ & 6.3--2000$\times$10$^{-13}$ & 1 & 7.8$\times$10$^{-29}$ & 9.0$\times$10$^{-30}$ & 9 \\
\ce{CH + H + M -> ^3CH2 + M} & 6.9$\times$10$^{-10}$ & 1.4--5.0$\times$10$^{-11}$ & 14 & 2.0$\times$10$^{-28}$ & & \\
\ce{CH + H + M -> ^1CH2 + M} & 2.1$\times$10$^{-10}$ & 1.4$\times$10$^{-11}$ & 15 & 4.4$\times$10$^{-31}$ & &\\
\ce{H + H + M -> H2 + M} & 1.9$\times$10$^{-10}$ & & & 1.7$\times$10$^{-33}$ & 4.0--250$\times$10$^{-33}$ & 2\\
\hline
\end{longtable*}

\subsection*{CRAHCN}

In Table~\ref{CRAHCN1}, we display the Lindemann coefficients for the three-body reactions in CRAHCN, along with their sources. When available, rate coefficients in CRAHCN are experimental values; However, the majority of rate coefficients have not been experimentally measured, and thus we use the consistently calculated values from this work. These three-body reactions, along with the one- and two-body reactions in Table~\ref{CRAHCN2}, make up CRAHCN.

\begin{longtable*}{clccc}
\caption{Lindemann coefficients for the three-body reactions in the consistent reduced atmospheric hybrid chemical network (CRAHCN), valid within the 50--400 K temperature range. Experimental values are used when available, and calculated rate coefficients from this work are used otherwise. $k_{\infty}$ is the second-order rate coefficient in the high pressure limit with units cm$^{3}$s$^{-1}$. $k_0$ is the third-order rate coefficient in the low pressure limit with units cm$^{6}$s$^{-1}$. These values fit into the pressure-dependent rate coefficient equation $k$ = $\frac{k_0 [M] / k_{\infty}}{1 + k_0 [M] / k_{\infty}} k_{\infty}$. \label{CRAHCN1}} \\
No. & Reaction equation & k$_\infty$(298) &  k$_0$(298) & Source(s) \\ \hline \\[-2mm]
*1. & \ce{HCN + CH + M -> HCNCH_a* + M* -> } & 2.7$\times$10$^{-10}$ & 3.3$\times$10$^{-29}$ & This work, \citet{Reference2044} \\
  & \ce{HCNCH_b* + M* -> CH2NC* + M* ->} & & & \\
  & \ce{CH2CN + M} & & & \\
2. & \ce{CN + CN + M -> NCCN + M} & 1.0$\times$10$^{-12}$ & 4.7$\times$10$^{-32}$ & \citet{Reference2045} \\
*3. & \ce{CN + CN + M -> CNCN + M} & 6.3$\times$10$^{-12}$ & 2.2$\times$10$^{-30}$ & This work \\
*4. & \ce{CN + NH + M -> HNCN + M} & 2.3$\times$10$^{-11}$ & 5.9$\times$10$^{-29}$ & This work\\
5. & \ce{CN + ^4N + M -> CN2 + M} & 1.1$\times$10$^{-10}$ & 2.8$\times$10$^{-32}$ & \citet{Reference2004}, \citet{Reference2003} \\
& & & & \citet{Reference2046} \\
*6. & \ce{CN + ^2N + M -> CN2 + M} & 1.6$\times$10$^{-10}$ & 1.1$\times$10$^{-29}$ & This work\\
*7. & \ce{CN + CH3 + M -> CH3CN + M} & 1.3$\times$10$^{-11}$ & 3.8$\times$10$^{-26}$ & This work \\
*8. & \ce{CN + ^3CH2 + M -> CH2CN + M} & 2.9$\times$10$^{-11}$ & 8.5$\times$10$^{-27}$ & This work\\
*9. & \ce{CN + ^1CH2 + M -> CH2CN + M} & 6.3$\times$10$^{-11}$ & 3.5$\times$10$^{-26}$ & This work \\
*10. & \ce{CN + CH + M -> ^3HCCN + M} & 1.1$\times$10$^{-11}$ & 6.5$\times$10$^{-28}$ & This work\\
*11. & \ce{CN + CH + M -> ^1HCCN + M -> } & 2.9$\times$10$^{-11}$ & 1.2$\times$10$^{-28}$ & This work \\
& \ce{^1HC2N + M} & & & \\
12. & \ce{CN + H + M -> HCN + M} & 4.5$\times$10$^{-11}$ & 1.7$\times$10$^{-30}$ & This work \\
*13. & \ce{^4N + ^3CH2 + M -> H2CN + M} & 1.0$\times$10$^{-10}$ & 1.2$\times$10$^{-28}$ & This work \\
14. & \ce{^4N + ^4N + M -> N2 + M} & 1.4$\times$10$^{-11}$ & 1.9$\times$10$^{-32}$ & This work, Average of experimental\\
15. & \ce{^4N + H + M -> NH + M} & 4.3$\times$10$^{-10}$ & 4.8$\times$10$^{-32}$ & This work, \citet{Reference2030}\\
*16. & \ce{^2N + ^3CH2 + M -> H2CN + M} & 2.1$\times$10$^{-10}$ & 6.2$\times$10$^{-28}$ & This work\\
*17. & \ce{^2N + ^1CH2 + M -> H2CN + M} & 2.4$\times$10$^{-10}$ & 1.5$\times$10$^{-27}$ & This work \\
*18. & \ce{^2N + ^2N + M -> N2 + M} & 5.7$\times$10$^{-11}$ & 1.7$\times$10$^{-33}$ & This work\\
*19. & \ce{^2N + H + M -> NH + M} & 7.7$\times$10$^{-10}$ & 7.0$\times$10$^{-32}$ & This work\\
*20. & \ce{CH4 + ^1CH2 + M -> C2H6 + M} & 7.1$\times$10$^{-11}$ & 7.2$\times$10$^{-24}$ & \citet{Reference509}, This work\\
21. & \ce{CH3 + CH3 + M -> C2H6 + M} & 6.0$\times$10$^{-11}$ & 2.5$\times$10$^{-26}$ & \citet{Reference451}, \citet{Reference2043},  \\
 & & & & \citet{Reference2042} \\
22. & \ce{CH3 + H + M -> CH4 + M} & 3.5$\times$10$^{-10}$ & 1.4$\times$10$^{-28}$ & \citet{Reference451}\\
*23. & \ce{^3CH2 + ^3CH2 + M -> C2H4 + M} & 5.3$\times$10$^{-11}$ & 2.6$\times$10$^{-26}$ & \citet{Reference451}, This work\\
24. & \ce{^3CH2 + H + M -> CH3 + M} & 2.0$\times$10$^{-10}$ & 4.8$\times$10$^{-29}$ & \citet{Reference451}, This work \\
*25. & \ce{^1CH2 + ^1CH2 + M -> C2H4 + M} & 7.1$\times$10$^{-11}$ & 5.4$\times$10$^{-25}$ & This work \\
*26. & \ce{^1CH2 + H2 + M -> CH4 + M} & 1.2$\times$10$^{-10}$ & 1.4$\times$10$^{-27}$ & \citet{Reference451}, This work \\
*27. & \ce{^1CH2 + H + M -> CH3 + M} & 2.3$\times$10$^{-10}$ & 3.7$\times$10$^{-28}$ & This work\\
*28. & \ce{CH + CH + M -> C2H2 + M} & 1.9$\times$10$^{-10}$ & 3.5$\times$10$^{-28}$ & \citet{Reference519}, \citet{Reference523}, This work\\
29. & \ce{CH + H2 + M -> CH3 + M} & 9.6$\times$10$^{-11}$ & 9.0$\times$10$^{-30}$ & Average of experimental values, \citet{Reference517} \\
*30. & \ce{CH + H + M -> ^3CH2 + M} & 3.2$\times$10$^{-11}$ & 2.0$\times$10$^{-28}$ & Average of experimental values, This work\\
*31. & \ce{CH + H + M -> ^1CH2 + M} & 1.4$\times$10$^{-11}$ & 4.4$\times$10$^{-31}$ & \citet{Reference2047}, This work \\
32. & \ce{H + H + M -> H2 + M} & 1.9$\times$10$^{-10}$ & 7.4$\times$10$^{-33}$ & This work, \citet{Reference451}\\
\hline
\multicolumn{5}{l}{\footnotesize $*$ Reactions with no previously known rate coefficients.} \\
\end{longtable*}

In Table~\ref{CRAHCN2}, we display the temperature-dependent Arrhenius parameters calculated for the one- and two-body reactions in CRAHCN.

\setlength\LTcapwidth{\textwidth}
\begin{longtable*}{clccccc}
\caption{Arrhenius coefficients for the one- and two-body reactions in the consistent reduced atmospheric hybrid chemical network (CRAHCN). Experimental values are used when available, and calculated rate coefficients from this work are used otherwise. For the reactions with barriers from this work, rate coefficients are calculated at 50, 100, 200, 298, and 400 K, and are fit to the modified Arrhenius expression $k(T) = \alpha \left(\frac{T}{300}\right)^{\beta} e^{-\gamma/T}$. Barrierless reaction rate coefficients do not typically vary by more than a factor of 1--3 for temperatures between 50 and 400 K \citep{Reference587,Reference495,Reference481,Reference577,Reference572}.   Intermediate molecules are labelled with a bullet, and are included to describe the precise reaction pathway for multi-step reactions. First- and second-order reactions with rate coefficients slower than k = 10$^{-21}$ cm$^{3}$s$^{-1}$ are not included in this network. First-order rate coefficients have units s$^{-1}$. Second-order rate coefficients have units cm$^{3}$s$^{-1}$. \label{CRAHCN2}} \\
No. & Reaction equation & Forw./Rev. & $\alpha$ &  $\beta$ & $\gamma$ & Source\\ \hline \\[-2mm]
33. & \ce{N2 + ^2N -> N2 + ^4N} & F & 5.4$\times$10$^{-12}$ & 0 & 1620 & \citet{Reference552} \\
34. & \ce{N2 + ^1CH2 -> N2 + ^3CH2} & F & 2.3$\times$10$^{-12}$ & -2.15 & 74 & \citet{Reference2050} \\
35. & \ce{H2CN + NH -> HCN + NH2} & F & 4.3$\times$10$^{-13}$ & 2.05 & 331 & This work \\
36. & \ce{H2CN + ^4N -> ^3H2CNN* -> N2 + ^3CH2} & F & 4.3$\times$10$^{-12}$ & 0 & 0 & This work \\
37. & \ce{H2CN + ^4N -> HCN + NH} & F & 1.0$\times$10$^{-10}$ & 0 & 200 & \citet{Reference573}\\
*38. & \ce{H2CN + ^2N -> ^3H2CNN* -> N2 + ^3CH2} & F & 3.2$\times$10$^{-11}$ & 0 & 0 & This work\\
*39. & \ce{H2CN + ^2N -> ^1H2CNN* -> N2 + ^1CH2} & F & 6.4$\times$10$^{-12}$ & 0 & 0 & This work\\
*40. & \ce{H2CN + CH -> ^1H2CNCH* -> } & F & 2.6$\times$10$^{-11}$ & -1.66 & 119 & This work\\
 & \ce{H2CNC_a* + H* -> H2CNC_b* + H* ->} & & & & & \\
 & \ce{CH2CN + H} & & & & & \\
*41. & \ce{H2CN + CH -> ^3H2CNCH* -> HCN + ^3CH2} & F & 2.2$\times$10$^{-11}$ & 0 & 0 & This work\\
*42. & \ce{H2CN + CH -> ^1H2CNCH* -> HCN + ^1CH2} & F & 6.4$\times$10$^{-10}$ & -2.04 & 904 & This work\\
43. & \ce{H2CN + H -> HCN + H2} & F & 2.2$\times$10$^{-11}$ & 0 & 0 & This work\\
44. & \ce{H2CN -> HCN + H} & F & 1.3$\times$10$^{+13}$ & 1.74 & 19060 & \citet{Reference598} + \\
 & & & & & & This work \\
45. & \ce{HCN + CN -> HNCCN* -> NCCN + H} & F & 1.5$\times$10$^{-13}$ & 0 & 0 & Avg. of exper. vals\\
46. & \ce{HCN + ^2N -> N2 + CH} & F & 6.8$\times$10$^{-11}$ & 0 & 0 & This work\\
*47. & \ce{HCN + ^1CH2 -> CH2HCN* -> } & F & 3.7$\times$10$^{-13}$ & 0 & 0 & This work\\
 & \ce{CH2NC_a* + H* -> CH2NC_b* + H* -> } & & & & & \\
 & \ce{CH2CN + H} & & & & & \\
48. & \ce{HCN + H -> HCNH* -> HNC + H} & F & 9.0$\times$10$^{-11}$ & 1.20 & 6249 & This work\\
49.  & \ce{HCN + H <- HCNH* <- HNC + H} & R & 8.9$\times$10$^{-11}$ & 0.80 & 649 & This work\\
50. & \ce{CN + NH -> HCN + ^4N} & F & 1.1$\times$10$^{-13}$ & 0.15 & 528 & This work\\
51. & \ce{CN + NH -> HNNC* -> CHN2_a* -> } & F & 4.6$\times$10$^{-12}$ & 1.22 & 263 & This work \\
 & \ce{CHN2_b* -> CH + N2} & & & & & \\
52. & \ce{CN + ^4N -> CN2$_{(\nu)}$* -> N2 + C} & F & 1.1$\times$10$^{-10}$ & 0 & 0 & Avg. of exper. vals \\
*53. & \ce{CN + ^2N -> CN2$_{(\nu)}$* -> N2 + C} & F & 1.6$\times$10$^{-10}$ & 0 & 0 & This work\\
54. & \ce{CN + CH4 -> HCN + CH3} & F & 1.5$\times$10$^{-11}$ & 0 & 940 & \citet{Reference451} \\
55. & \ce{CN + CH3 -> HCN + ^3CH2} & F & 6.7$\times$10$^{-12}$ & 0 & 0 & This work\\
56. & \ce{CN + CH -> HCN + C} & F & 1.4$\times$10$^{-11}$ & 0 & 0 & This work\\
57. & \ce{CN + CH -> HNC + C} & F & 5.4$\times$10$^{-13}$ & 0 & 0 & This work\\
58. & \ce{CN + H2 -> HCN + H} & F & 4.1$\times$10$^{-12}$ & 1.55 & 1510 & \citet{Reference1025} \\
59. & \ce{NH + NH -> N2H2* -> N2H* + H* ->} & F & 3.5$\times$10$^{-12}$ & 0 & 0 & \citet{Reference2021} \\
& \ce{N2 + H + H} & & & & & \\
60. & \ce{NH + NH -> NH2 + ^4N} & F & 1.9$\times$10$^{-12}$ & 0.47 & 3738 & This work\\
61. & \ce{NH + ^4N -> N2H* -> N2 + H} & F & 2.5$\times$10$^{-11}$ & 0 & 0 & \citet{Reference525}\\
62. & \ce{NH + ^2N -> N2H* -> N2 + H} & F & 8.8$\times$10$^{-11}$ & 0 & 0 & \citet{Reference598} + \\
 & & & & & & This work \\
63. & \ce{NH + CH3 -> H3CNH* -> ^1H2CNH* + H* ->} & F & 2.0$\times$10$^{-11}$ & 0 & 0 & This work\\
 & \ce{H2CN + H2} & & & & & \\
64. & \ce{NH + ^3CH2 -> ^1H2CNH* -> H2CN + H} & F & 1.9$\times$10$^{-11}$ & 0 & 0 & This work \\
*65. & \ce{NH + ^1CH2 -> ^3H2CNH* -> H2CN + H} & F & 4.6$\times$10$^{-11}$ & 0 & 0 & This work\\
66. & \ce{NH + CH -> HNCH* -> HCN + H} & F & 4.5$\times$10$^{-11}$ & 0 & 0 & This work\\
67. & \ce{NH + CH -> HNCH* -> HNC + H} & F & 5.2$\times$10$^{-11}$ & 1.04 & 2551  & This work\\
68. & \ce{NH + H -> H2 + ^4N} & F & 2.2$\times$10$^{-12}$ & 1.55 & 103 & \citet{Reference527}\\
69. & \ce{^4N + CH3 -> ^3H3CN* -> H2CN + H} & F & 7.7$\times$10$^{-11}$ & 0 & 0 & \citet{Reference442} + \\
& & & & & & \citet{Reference444}\\
70. & \ce{^4N + ^3CH2 -> H2CN$_{(\nu)}$* -> HCN + H} & F & 1.0$\times$10$^{-10}$ & 0 & 0 & \citet{Reference598} + \\
 & & & & & & This work \\
71. & \ce{^4N + ^1CH2 -> ^4H2CN -> ^3HCN* + H* ->} & F & 1.9$\times$10$^{-10}$ & 0 & 0 & \citet{Reference598} + \\
 & \ce{CN + H + H} & & & & & This work \\
72. & \ce{^4N + CH -> ^3HCN* -> CN + H} & F & 1.1$\times$10$^{-10}$ & 0 & 0 & Avg. of exper. vals\\
73. & \ce{^2N + CH4 -> H3CNH* -> ^1H2CNH* + H* -> } & F & 4.8$\times$10$^{-11}$ & 0 & 750 & \citet{Reference549}\\
 & \ce{H2CN + H2} & & & & & \\
74. & \ce{^2N + CH3 -> ^{3,1}H3CN* -> ^{3,1}H2CNH*} & F & 2.3$\times$10$^{-10}$ & 0 & 0 & \citet{Reference598} + \\
 & \ce{H2CN + H} & & & & & This work \\
75. & \ce{^2N + ^3CH2 -> H2CN$_{(\nu)}$* -> HCN + H} & F & 2.1$\times$10$^{-10}$ & 0 & 0 & \citet{Reference598} + \\
 & & & & & & This work \\
76. & \ce{^2N + ^3CH2 -> ^4H2CN* -> ^3HCN* + H* -> } & F & 3.5$\times$10$^{-10}$ & 0 & 0 & \citet{Reference598} + \\
 & \ce{CN + H + H} & & & & & This work \\
77. & \ce{^2N + ^1CH2 -> H2CN$_{(\nu)}$* -> HCN + H} & F & 2.4$\times$10$^{-10}$ & 0 & 0 & \citet{Reference598} + \\
 & & & & & & This work \\
78. & \ce{^2N + CH -> ^3HCN* -> CN + H} & F & 3.5$\times$10$^{-10}$ & 0 & 0 & \citet{Reference598} + \\
 & & & & & & This work \\
79. & \ce{^2N + H2 -> NH2$_{(\nu)}$* -> NH + H} & F & 4.2$\times$10$^{-11}$ & 0 & 880 & \citet{Reference549}\\
80. & \ce{CH4 + ^3CH2 -> CH3 + CH3} & F & 5.4$\times$10$^{-13}$ & 7.45 & 3401 & \citet{Reference598} + \\
 & & & & & & This work \\
81. & \ce{CH4 + ^3CH2 <- CH3 + CH3} & R & 5.3$\times$10$^{-11}$ & 0 & 0 & \citet{Reference598} + \\
 & & & & & & This work \\
82. & \ce{CH4 + ^1CH2 -> C2H6$_{(\nu)}$* -> CH3 + CH3} & F & 7.1$\times$10$^{-11}$ & 0 & 0 & \citet{Reference509}\\
83. & \ce{CH4 + CH -> CH4-CH* -> C2H5* -> } &  F & 9.8$\times$10$^{-11}$ & 0 & 0 & \citet{Reference451}\\
& \ce{C2H4 + H} & & & & & \\
84. & \ce{CH4 + H -> CH3 + H2} & F & 5.9$\times$10$^{-13}$ & 3.0 & 4045 & \citet{Reference451}\\
85. & \ce{CH4 + H <- CH3 + H2} & R & 7.0$\times$10$^{-14}$ & 2.74 & 4740 & \citet{Reference451}\\
86. & \ce{CH3 + ^3CH2 -> C2H5* -> C2H4 + H} & F & 7.0$\times$10$^{-11}$ & 0 & 0 & \citet{Reference451}\\
87. & \ce{CH3 + ^1CH2 -> C2H5* -> C2H4 + H} & F & 1.3$\times$10$^{-10}$ & 0 & 0 & \citet{Reference598} + \\
 & & & & & & This work \\
88. & \ce{CH3 + CH -> ^1CH3CH* -> C2H3* + H* ->} & F & 5.5$\times$10$^{-11}$ & 0 & 0 & This work\\
 & \ce{C2H4} & & & & &\\
89. & \ce{CH3 + CH -> ^3CH3CH* -> C2H3* + H* ->} & F & 2.4$\times$10$^{-10}$ & 0 & 0 & This work\\
 & \ce{C2H2 + H + H} & & & & & \\
90. & \ce{CH3 + H -> ^3CH2 + H2} & F & 2.6$\times$10$^{-11}$ & 1.15 & 6529 & \citet{Reference598} + \\
 & & & & & & This work \\
91. & \ce{CH3 + H <- ^3CH2 + H2} & R & 8.1$\times$10$^{-14}$ & 9.04 & 1450 & \citet{Reference598} + \\
 & & & & & & This work \\
92. & \ce{^3CH2 + ^3CH2 -> C2H4$_{(\nu)}$* -> C2H2 + H2} & F & 5.3$\times$10$^{-11}$ & 0 & 0 & \citet{Reference451}\\
93. & \ce{^3CH2 + ^1CH2 -> ^3C2H4* -> C2H3* + H -> } & F & 2.1$\times$10$^{-10}$ & 0 & 0 & \citet{Reference598} + \\
 & \ce{C2H2 + H + H} & & & & & This work \\
94. & \ce{^3CH2 + CH -> C2H3* -> C2H2 + H} & F & 6.3$\times$10$^{-11}$ & 0 & 0 & This work\\
*95. & \ce{^3CH2 + CH -> ^4C2H3* -> ^3C2H2* + H* ->} & F & 1.1$\times$10$^{-10}$ & 0 & 0 & This work\\
& \ce{C2H + H + H} & & & & & \\
96. & \ce{^3CH2 + H -> CH3$_{(\nu)}$* -> CH + H2} & F & 2.0$\times$10$^{-10}$ & 0 & 0 & \citet{Reference451} \\
97. & \ce{^1CH2 + ^1CH2 -> C2H4$_{(\nu)}$* -> C2H2 + H2} & F & 7.1$\times$10$^{-11}$ & 0 & 0 & \citet{Reference598} + \\
 & & & & & & This work \\
*98. & \ce{^1CH2 + CH -> C2H3* -> C2H2 + H} & F & 1.6$\times$10$^{-10}$ & 0 & 0 & This work\\
99. & \ce{^1CH2 + H2 -> CH4$_{(\nu)}$* -> CH3 + H} & F & 1.2$\times$10$^{-10}$ & 0 & 0 & \citet{Reference451}\\
100. & \ce{^1CH2 + H -> CH3$_{(\nu)}$* -> CH + H2} & F & 1.8$\times$10$^{-10}$ & 0 & 0 & This work + \\
& & & & & & \citet{Reference2040}\\
*101. & \ce{^1CH2 + H -> CH3$_{(\nu)}$* -> ^3CH2 + H} & F & 4.6$\times$10$^{-11}$ & 0 & 0 & This work +\\
& & & & & &  \citet{Reference2040}\\
102. & \ce{CH + CH -> ^3C2H2* -> C2H + H} & F & 1.4$\times$10$^{-10}$ & 0 & 0 & \citet{Reference598} + \\
 & & & & & & This work \\
103. & \ce{CH + H2 -> CH3$_{(\nu)}$* -> ^3CH2 + H} & F & 3.1$\times$10$^{-10}$ & 0 & 1650 & \citet{Reference575}\\
104. & \ce{CH + H -> ^3CH2$_{(\nu)}$* -> C + H2} & F & 6.9$\times$10$^{-10}$ & 0 & 0 & \citet{Reference598} + \\
 & & & & & & This work \\
\hline
\multicolumn{7}{l}{\footnotesize $*$ Reactions with no previously known rate coefficients.} \\
\multicolumn{7}{l}{\footnotesize Vibrationally excited molecules are labeled with a $_{(\nu)}$ subscript.} \\
\end{longtable*}

\subsection*{Experimental Data}

In Table~\ref{TableExp}, we list all experimental rate coefficients for the reactions calculated in this work.

\setlength\LTcapwidth{\textwidth}
\begin{longtable*}{ccccc}
\caption{Available experimental or recommended reaction rate coefficients for the new reactions in this paper. For brevity, only the 10 most recent measurements are included; for a complete listing, we refer the reader to the NIST Chemical Kinetics Database \citep{Reference494}. First-order rate coefficients have units s$^{-1}$. Second-order rate coefficients have units cm$^{3}$s$^{-1}$. Third-order rate coefficients have units cm$^{6}$s$^{-1}$.\label{TableExp}} \\
k(298K) & Technique & Temp. (K) & Pressure (Torr) & Reference(s)\\ \hline \\[-2mm]
\ce{H2CN + ^4N -> HCN + NH} & & & & \\
4.4$\times$10$^{-11}$ & M & 298 & 1 &	\citet{Reference573}\\
\ce{H2CN + H -> HCN + H2} & & & & \\
8.3$\times$10$^{-11}$ & Z & independent & &	\citet{Reference446}\\
\ce{HCN + CN -> NCCN + H} & & & & \\
4.1$\times$10$^{-13}$ & M & 298 & 20 & \citet{Reference2000} \\
3.2$\times$10$^{-14}$ & M & 298 & 100--600 & \citet{Reference2002} \\
1.8$\times$10$^{-14}$ & M & 300 & 5--25 & \citet{Reference2001} \\
3.2$\times$10$^{-14}$ & S & 298 &  & \citet{Reference1025} \\
\ce{HCN + CH -> products} & & & & \\
2.7$\times$10$^{-10}$ & M & 298 & 100 & \citet{Reference2044} \\
\ce{CN + CN + M -> NCCN + M} & & & & \\
4.7$\times$10$^{-32}$ & M & 298 & 1--27 & \citet{Reference2045} \\
3.3--4.9$\times$10$^{-29}$ & S & 298 &  & \citet{Reference1025} \\
\ce{CN + CN -> NCCN} & & & & \\
1.0$\times$10$^{-12}$ & M & 298 & 1--27 & \citet{Reference2045} \\
9.4$\times$10$^{-12}$ & S & 298 &  & \citet{Reference1025} \\
\ce{CN + ^4N -> N2 + C} & & & & \\
1.2$\times$10$^{-10}$ & M & 298 & 7 & \citet{Reference2004} \\
1.0$\times$10$^{-10}$ & M & 300 & 2 & \citet{Reference2003} \\
3.0$\times$10$^{-10}$ & S & 298 &  & \citet{Reference451} \\
\ce{CN + ^4N + M -> CN2 + M} & & & & \\
2.8$\times$10$^{-32}$ & M & 298 & 1--10 & \citet{Reference2046} \\
\ce{CN + CH4 -> HCN + CH3} & & & & \\
1.1$\times$10$^{-12}$ & M & 298 &  & \citet{Reference2011} \\
1.1$\times$10$^{-12}$ & M & 295 & 53 & \citet{Reference2013} \\
9.2$\times$10$^{-13}$ & M & 298 & 4--38 & \citet{Reference2009} \\
7.8$\times$10$^{-13}$ & M & 298 & 1--200 & \citet{Reference2012} \\
7.5$\times$10$^{-13}$ & M & 297 & 20 & \citet{Reference2008} \\
7.4$\times$10$^{-13}$ & M & 298 & & \citet{Reference2006} \\
7.4$\times$10$^{-13}$ & M & 298 & & \citet{Reference2007} \\
5.8$\times$10$^{-13}$ & M & 298 & 47--490 & \citet{Reference2005} \\
5.6$\times$10$^{-13}$ & M & 294 & 5--30 & \citet{Reference2010} \\
6.4$\times$10$^{-13}$ & S & 298 &  & \citet{Reference451} \\
\ce{CN + H2 -> HCN + H} & & & & \\
4.9$\times$10$^{-14}$ & M & 294 & 52 & \citet{Reference2013} \\
2.5--2.9$\times$10$^{-14}$ & M & 298 & 50 & \citet{Reference2014} \\
2.5--2.7$\times$10$^{-14}$ & M & 294 & 3--6 & \citet{Reference2015} \\
2.7$\times$10$^{-14}$ & M & 298 & 50--500 & \citet{Reference2016} \\
2.7$\times$10$^{-14}$ & M & 298 & 50 & \citet{Reference2019} \\
2.2--2.6$\times$10$^{-14}$ & M & 298 & 1--200 & \citet{Reference2012} \\
2.5$\times$10$^{-14}$ & M & 295 & 10--30 & \citet{Reference2020} \\
1.4$\times$10$^{-14}$ & M & 298 & 3--50  & \citet{Reference2018} \\
1.2$\times$10$^{-14}$ & M & 298 & 4--21 & \citet{Reference2017} \\
2.6$\times$10$^{-14}$ & S & 298 &  & \citet{Reference1025} \\
\ce{CN + H + M -> HCN + M} & & & & \\
1.3--1.6$\times$10$^{-30}$ & S & 298 &  & \citet{Reference1025} \\
\ce{NH + NH -> N2H2* -> products} & & & & \\
3.5$\times$10$^{-12}$ & M & 298 & 3--10 & \citet{Reference2021} \\
\ce{NH + ^4N -> N2 + H} & & & & \\
2.5$\times$10$^{-11}$ & M & 298 & 11--15 & \citet{Reference525}\\
2.6$\times$10$^{-11}$ & S & 300 & & \citet{Reference526}\\
\ce{NH + H -> H2 + ^4N} & & & & \\
3.2$\times$10$^{-12}$ & M & 298 & 2--8 &\citet{Reference527}\\
\ce{^4N + ^4N + M -> N2 + M} & & & & \\
1.3$\times$10$^{-31}$ & M & 298 & & \citet{Reference2023} \\
1.1--3.2$\times$10$^{-32}$ & M & 300 & & \citet{Reference2025} \\
1.1--2.3$\times$10$^{-32}$ & M & 300 & & \citet{Reference2024} \\
2.3$\times$10$^{-32}$ & M & 298 & 0.3--7 & \citet{Reference2029} \\
1.7$\times$10$^{-32}$ & M & 298 & 0.5--1.3 & \citet{Reference2022} \\
7.4--9.7$\times$10$^{-33}$ & M & 298 & 1--13 & \citet{Reference2027} \\
6.1--7.2$\times$10$^{-33}$ & M & 298 & 2--10 & \citet{Reference1056} \\
3.8--5.3$\times$10$^{-33}$ & M & 298 & 2--10 & \citet{Reference2028} \\
3.3$\times$10$^{-33}$ & M & 297 & 14 & \citet{Reference2026} \\
2.2$\times$10$^{-33}$ & M & 298 & 2--9 & \citet{Reference1057} \\
\ce{^4N + CH3 -> HCN + H2} & & & & \\
8.6$\times$10$^{-12}$ & M & 298 & 0.3--1.6 &	\citet{Reference442},\\
& & & & \citet{Reference444}\\
\ce{^4N + CH3 -> H2CN + H} & & & & \\
7.7$\times$10$^{-11}$ & M & 298 & 0.3--1.6 &	\citet{Reference442},\\
& & & & \citet{Reference444}\\
5.0$\times$10$^{-11}$ & S & independent & & \citet{Reference445}\\
\ce{^4N + CH -> ^3HCN -> CN + H} & & & & \\
1.6$\times$10$^{-10}$ & M & 298 & 4 &\citet{Reference570}\\
1.2--1.4$\times$10$^{-10}$ & M & 296 & 5 &\citet{Reference572}\\
2.1$\times$10$^{-11}$ & M & 298 & 5--15 &\citet{Reference571}\\
\ce{^4N + H + M -> NH + M} & & & & \\
3.1--6.4$\times$10$^{-32}$ & M & 298 &  & \citet{Reference2030} \\
$>$1.3$\times$10$^{-33}$ & M & 298 & 2.5--4.5 & \citet{Reference2031} \\
\ce{^2N + CH4 -> H2CNH + H} & & & & \\
$^a$4.5$\times$10$^{-12}$ & M & 298 & 700 & \citet{Reference579},\\
& & & & \citet{Reference580}\\
$^a$3.7$\times$10$^{-12}$ & M & 300 & 6 & \citet{Reference555},\\
& & & & \citet{Reference580}\\
$^a$2.7$\times$10$^{-12}$ & M & 295 & 20 & \citet{Reference581},\\
& & & & \citet{Reference580}\\
$^a$2.4$\times$10$^{-12}$ & M & 300 & 3--5 & \citet{Reference559},\\
& & & & \citet{Reference580}\\
$^a$3.2$\times$10$^{-12}$ & S & 298 & & \citet{Reference549},\\
& & & & \citet{Reference580}\\
\ce{^2N + H2 -> NH2* -> NH + H} & & & & \\
5.0$\times$10$^{-12}$ & M & 300 & 3--5 &\citet{Reference559}\\
3.5$\times$10$^{-12}$ & M & 300 & 6 &\citet{Reference555}\\
2.7$\times$10$^{-12}$ & M & 300 & 2--5 &\citet{Reference556}\\
2.4$\times$10$^{-12}$ & M & 300 & 753 &\citet{Reference552}\\
2.3$\times$10$^{-12}$ & M & 295 & 30 &\citet{Reference548}\\
2.3$\times$10$^{-12}$ & M & 300 & 1--3 &\citet{Reference554}\\
2.1$\times$10$^{-12}$ & M & 300 & 26 &\citet{Reference557}\\
1.8$\times$10$^{-12}$ & M & 298 & 1 &\citet{Reference553}\\
1.7$\times$10$^{-12}$ & M & 300 & 50 &\citet{Reference558}\\
2.2$\times$10$^{-12}$ & S & 200--300 &  &\citet{Reference549}\\
\ce{CH4 + ^3CH2 -> 2CH3} & & & & \\
$<$5.0$\times$10$^{-14}$ & M & 298 & 10 &\citet{Reference510}\\
$^f$3.1$\times$10$^{-19}$ & M & 298 & 2--3 &\citet{Reference568}\\
$<$3.0$\times$10$^{-19}$ & S & 298 & &\citet{Reference509}\\
\ce{CH4 + ^1CH2 -> 2CH3} & & & & \\
7.3$\times$10$^{-11}$ & M & 298 & 10$^{-4}$--10 &\citet{Reference515}\\
7.0$\times$10$^{-11}$ & M & 295 & 6 &\citet{Reference514}\\
1.9$\times$10$^{-12}$ & M & 298 & 5--20 &\citet{Reference510}\\
7.1$\times$10$^{-11}$ & S & & &\citet{Reference509}\\
\ce{CH4 + CH -> C2H4 + H} & & & & \\
3.0$\times$10$^{-10}$ & M & 298 & 30--100 & \citet{Reference521}\\
1.0$\times$10$^{-10}$ & M & 298 & 100 & \citet{Reference562}\\
9.8$\times$10$^{-11}$ & M & 298 & 100 & \citet{Reference561}\\
9.1$\times$10$^{-11}$ & M & 298 & 50--300 & \citet{Reference563}\\
8.9$\times$10$^{-11}$ & M & 295 & 9--12 & \citet{1997AA...323..644C}\\
6.7$\times$10$^{-11}$ & M & 298 & 100 & \citet{Reference566}\\
3.3$\times$10$^{-11}$ & M & & &\citet{Reference520}\\
2.5$\times$10$^{-12}$ & M & 298 & 100 &\citet{Reference519}\\
2.0$\times$10$^{-12}$ & M & 298 & 1-500 &\citet{Reference523}\\
9.8$\times$10$^{-11}$ & S & 298 & &\citet{Reference451}\\
\ce{CH4 + H -> CH3 + H2} & & & & \\
3.5$\times$10$^{-17}$ & M & 298 &  & \citet{Reference452}\\
1.7$\times$10$^{-17}$ & M & 298 & 0.55 & \citet{Reference450}\\
8.2$\times$10$^{-19}$ & S & 300 & &\citet{Reference451}\\
\ce{CH3 + CH3 + M -> C2H6 + M} & & & & \\
3.3$\times$10$^{-26}$ & M & 298 & 5--600 & \citet{Reference2043} \\
1.7$\times$10$^{-26}$ & M & 298 & 5--500 & \citet{Reference2042} \\
\ce{2CH3 -> C2H6} & & & & \\
6.5$\times$10$^{-11}$ & M & 300 & high-pressure limit &\citet{Reference484}\\
6.5$\times$10$^{-11}$ & M & 298 & high-pressure limit &\citet{Reference493}\\
6.0$\times$10$^{-11}$ & M & 298 & high-pressure limit &\citet{Reference483}\\
6.0$\times$10$^{-11}$ & M & 298 & high-pressure limit &\citet{Reference489}\\
$^d$5.9$\times$10$^{-11}$ & M & 292 & 758 &\citet{Reference486}\\
5.8$\times$10$^{-11}$ & M & 298 & 750 &\citet{Reference490}\\
5.5$\times$10$^{-11}$ & M & 298 & high-pressure limit &\citet{Reference487}\\
5.2$\times$10$^{-11}$ & M & 298 & 100 &\citet{Reference485}\\
4.6$\times$10$^{-11}$ & M & 300 & 1 &\citet{Reference513}\\
4.0$\times$10$^{-11}$ & M & 302 & 81--571 &\citet{Reference492}\\
3.5$\times$10$^{-11}$ & M & 308 & 86 &\citet{Reference491}\\
6.0$\times$10$^{-11}$ & S & independent & high-pressure limit &\citet{Reference451}\\
4.4$\times$10$^{-11}$ & S & 298 & high-pressure limit &\citet{Reference472}\\
\ce{CH3 + ^3CH2 -> C2H5* -> C2H4 + H} & & & & \\
2.1$\times$10$^{-10}$ & M & 300 & 1 l&\citet{Reference513}\\
1.1$\times$10$^{-10}$ & M & 298 & 1 &\citet{Reference578}\\
1.0$\times$10$^{-10}$ & M & 308 & 50--700 &\citet{Reference512}\\
5.0$\times$10$^{-11}$ & M & & 200 &\citet{Reference511}\\
7.0$\times$10$^{-11}$ & S & independent & &\citet{Reference451}\\
7.0$\times$10$^{-11}$ & S & 298 & &\citet{Reference509}\\
\ce{CH3 + ^1CH2 -> C2H4 + H} & & & & \\
3.0$\times$10$^{-11}$ & S & & &\citet{Reference509}\\
\ce{CH3 + H2 -> CH4 + H} & & & & \\
$^c$1.3$\times$10$^{-20}$ & M & 300 &  &	\citet{Reference588}\\
1.2$\times$10$^{-20}$ & S & 300 &  &	\citet{Reference509}\\
9.6$\times$10$^{-21}$ & S & 300 &  &	\citet{Reference451}\\
\ce{CH3 + H + M -> CH4 + M} & & & & \\
1.5--55$\times$10$^{-29}$ & M & 295	 & 6--15 & \citet{Reference2033} \\
2.6$\times$10$^{-29}$ & M & 298 & 2--10 & \citet{Reference2032} \\
6.2--30$\times$10$^{-29}$ & S & 298 &  & \citet{Reference451} \\
\ce{CH3 + H -> CH4} & & & & \\
4.7$\times$10$^{-10}$ & M & 300 & high-pressure limit &\citet{Reference468}\\
3.3$\times$10$^{-10}$ & M & 308 & high-pressure limit &\citet{Reference470}\\
2.5$\times$10$^{-10}$ & M & 308 & 300 &\citet{Reference471}\\
2.0$\times$10$^{-10}$ & M & 296 & 735--755 &\citet{Reference473}\\
1.5$\times$10$^{-10}$ & M & 300 & high-pressure limit &\citet{Reference474}\\
3.4$\times$10$^{-10}$ & F & 298 & high-pressure limit &\citet{Reference475}\\
3.5$\times$10$^{-10}$ & S & independent & high-pressure limit &\citet{Reference469}\\
3.5$\times$10$^{-10}$ & S & independent & high-pressure limit &\citet{Reference451}\\
2.0$\times$10$^{-10}$ & S &  298 & high-pressure limit &\citet{Reference472}\\
\ce{^3CH2 + ^3CH2 -> C2H2 + 2H} & & & & \\
5.3$\times$10$^{-11}$ & M & 298 & 20--700 &\citet{Reference510}\\
5.3$\times$10$^{-11}$ & S & 300 & &\citet{Reference451}\\
\ce{^3CH2 + ^1CH2 -> C2H2 + H2} & & & & \\
3.0$\times$10$^{-11}$ & S & & &\citet{Reference509}\\
\ce{^3CH2 + H2 -> CH3 + H} & & & & \\
$<$5.0$\times$10$^{-14}$ & M & 298 & 10 &\citet{Reference510}\\
$<$6.9$\times$10$^{-15}$ & M & 295 & 8 & \citet{Reference545}\\
$<$5.0$\times$10$^{-15}$ & M & & 10 &\citet{Reference544}\\
\ce{^3CH2 + H -> CH3* -> CH + H2} & & & & \\
2.7$\times$10$^{-10}$ & M & 285 & 2 &\citet{Reference504}\\
2.7$\times$10$^{-10}$ & M & 298 & 2 &\citet{Reference507}\\
$^e$2.6$\times$10$^{-10}$ & M & 300 & 2 &\citet{Reference503}\\
1.8$\times$10$^{-10}$ & M & 298 & 1--2 &\citet{Reference505}\\
$^c$1.4$\times$10$^{-10}$ & M & 300 & 100 &\citet{Reference506}\\
8.3$\times$10$^{-11}$ & M & 298 & 2 &\citet{Reference508}\\
2.0$\times$10$^{-10}$ & S & 300 & &\citet{Reference451}\\
2.7$\times$10$^{-10}$ & S & 298 & &\citet{Reference509}\\
\ce{^1CH2 + ^1CH2 -> C2H2 + 2H} & & & & \\
5.0$\times$10$^{-11}$ & S & & &\citet{Reference509}\\
\ce{^1CH2 + H2 -> CH4* -> CH3 + H} & & & & \\
1.3$\times$10$^{-10}$ & M & 295 & 6 &\citet{Reference514}\\
1.1$\times$10$^{-10}$ & M & 298 & 10$^{-4}$--10 &\citet{Reference515}\\
7.0$\times$10$^{-12}$ & M & 298 & 10 &\citet{Reference510}\\
1.2$\times$10$^{-10}$ & S &  & &\citet{Reference509}\\
1.2$\times$10$^{-10}$ & S & independent & &\citet{Reference451}\\
\ce{^1CH2 + H -> CH + H2} & & & & \\
5.0$\times$10$^{-11}$ & S & & &\citet{Reference509}\\
\ce{2CH -> C2H2} & & & & \\
2.0$\times$10$^{-10}$ & M & 298 & 1--330 &\citet{Reference519}\\
1.7$\times$10$^{-10}$ & M & 298 & 1--500 &\citet{Reference523}\\
\ce{CH + H2 + M -> CH3 + M} & & & & \\
9.0$\times$10$^{-30}$ & M & 298 & 2--591 & \citet{Reference517} \\
\ce{CH + H2 -> CH3} & & & & \\
1.6$\times$10$^{-10}$ & M & 294 & high-pressure limit &\citet{Reference575}\\
5.1$\times$10$^{-11}$ & M & 300 & 750 &\citet{Reference516}\\
4.5$\times$10$^{-11}$ & M & 298 & 591 &\citet{Reference517}\\
4.5$\times$10$^{-11}$ & M & 279 & 600 &\citet{Reference518}\\
3.0$\times$10$^{-11}$ & M & 294 & 750 &\citet{Reference576}\\
2.3$\times$10$^{-11}$ & M & 298 & 100 &\citet{Reference521}\\
1.7$\times$10$^{-11}$ & M & & &\citet{Reference520}\\
1.0$\times$10$^{-12}$ & M & 298 & 1--9 &\citet{Reference519}\\
\ce{CH + H2 -> CH3* -> ^3CH2 + H} & & & & \\
$^c$9.1$\times$10$^{-13}$ & M & 300 & 100 &\citet{Reference506}\\
1.2$\times$10$^{-12}$ & M & 294 & 400 &\citet{Reference575}\\
4.5$\times$10$^{-11}$ & M & 298 & 591 & \citet{Reference517}\\
\ce{CH + H -> C + H2} & & & & \\
5.0$\times$10$^{-11}$ & M & 298 &  & \citet{Reference2049} \\
\ce{CH + H -> products} & & & & \\
1.4$\times$10$^{-11}$ & M & 297 & 2 & \citet{Reference2047} \\
\ce{H + H + M -> H2 + M} & & & & \\
2.5$\times$10$^{-31}$ & M & 303 &  & \citet{Reference2039} \\
7.0--20$\times$10$^{-33}$ & M & 295 & 6 & \citet{Reference2036} \\
5.9--10$\times$10$^{-33}$ & M & 298 & 500-1500 & \citet{Reference2035} \\
9.4$\times$10$^{-33}$ & M & 298 & 50--250 & \citet{Reference1049} \\
7.0--9.2$\times$10$^{-33}$ & M & 298 & 2--15 & \citet{Reference2037} \\
5.6--6.7$\times$10$^{-33}$ & M & 297 & 6--18 & \citet{Reference2034} \\
4.0--5.1$\times$10$^{-33}$ & M & 298 & 1.5--4.5 & \citet{Reference2038} \\
6.4--92$\times$10$^{-33}$ & S & 298 &  & \citet{Reference1085} \\
9.1$\times$10$^{-33}$ & S & 298 &  & \citet{Reference509} \\
6.0--8.8$\times$10$^{-33}$ & S & 298 &  & \citet{Reference451} \\
\hline
\multicolumn{5}{l}{\footnotesize M: Monitoring decay of reactants and/or production of products.} \\
\multicolumn{5}{l}{\footnotesize Z: Zero activation energy value. Calculated by numerical modeling using the chemical compositions of the flames of CH$_4$ + air.} \\
\multicolumn{5}{l}{\footnotesize S: Suggested value based on experiments and/or evaluations at a range of temperatures.} \\
\end{longtable*}

\setlength\LTcapwidth{\textwidth}
\begin{longtable*}{lccccc}
\caption{Lennard-Jones force constants used in this study. Values are obtained from viscosity data when possible. \label{TableLJ}} \\
Molecule & $\sigma$ ($\AA$) & $\epsilon/k_b$ (K) & Source \\ \hline \\[-2mm]
\ce{NCCN} & 4.38 & 339 & \citet{Reference2056} \\
\ce{CNCN} & $^a$4.38 & $^a$339 & \citet{Reference2056} \\
\ce{CH3CN} & $^b$4.418 & $^b$230 & \citet{Reference2056} \\
\ce{CH2CN} &$^c$4.232 & $^c$205 & \citet{Reference2056} \\
\ce{HCNCH_a} & $^c$4.232 & $^c$205 & \citet{Reference2056} \\
\ce{HNCN} & $^d$3.996 & $^d$190 & \citet{Reference2056} \\
\ce{CN2} & $^d$3.996 & $^d$190 & \citet{Reference2056} \\
\ce{^3HCCN} & $^e$4.221 & $^e$185 & \citet{Reference2056} \\
\ce{^1HCCN} & $^e$4.221 & $^e$185 & \citet{Reference2056} \\
\ce{H2CN} & $^f$3.63 & $^f$569.1 & \citet{Reference2057} \\
\ce{HCN} & 3.63 & 569.1 & \citet{Reference2057} \\
\ce{N2} & 3.681 & 91.5  & \citet{Reference2056}\\
\ce{NH} & $^g$2.75 & $^g$80 & \citet{jetsurf2.0}\\
\ce{C2H6} & 4.418 & 230 & \citet{Reference2056} \\
\ce{C2H4} & 4.232 & 205 & \citet{Reference2056} \\
\ce{C2H2} & 4.221 & 185 & \citet{Reference2056} \\
\ce{CH4} & 3.822 & 136.5 & \citet{Reference2056}\\
\ce{CH3} & 3.8 & 144 & \citet{jetsurf2.0}\\
\ce{^3CH2} & 3.8 & 144 & \citet{jetsurf2.0}\\
\ce{^1CH2} & 3.8 & 144 & \citet{jetsurf2.0}\\
\ce{H2} & 2.968 & 33.3  & \citet{Reference2056}\\
\hline
\multicolumn{4}{l}{\footnotesize $^a$ L-J parameters based on those for \ce{NCCN}} \\
\multicolumn{4}{l}{\footnotesize $^b$ L-J parameters based on those for \ce{C2H6}} \\
\multicolumn{4}{l}{\footnotesize $^c$ L-J parameters based on those for \ce{C2H4}} \\
\multicolumn{4}{l}{\footnotesize $^d$ L-J parameters based on those for \ce{CO2}} \\
\multicolumn{4}{l}{\footnotesize $^e$ L-J parameters based on those for \ce{C2H2}} \\
\multicolumn{4}{l}{\footnotesize $^f$ L-J parameters based on those for \ce{HCN}} \\
\multicolumn{4}{l}{\footnotesize $^g$ L-J parameters based on those for \ce{OH} and \ce{CH}} \\
\end{longtable*}

\subsection*{Theoretical Case Studies}

The following theoretical case studies provide additional details for some of the non-standard reactions in CRAHCN. For example, these reactions might have an excited intermediate, or may have a barrier that isn't detected by our chosen computational method.

\subsubsection*{Case Study 1: \ce{H2CN + ^4N -> HCN + NH}}

One experiment has measured the rate coefficient for this reaction to have a value of $k$(298 K) = 4.4$\times$10$^{-11}$ cm$^{3}$s$^{-1}$ \citep{Reference573}. No isotope effect was observed, which is consistent with a barrierless reaction. Using three temperature data points (200K, 298K, 363K), the authors suggest Arrhenius parameters indicative of a very small barrier. However, with the small number of data points, and the data uncertainties, complete temperature independence of this reaction would also fit these data points \citep{Reference573}.

\citet{Reference573} suggest this reaction either proceeds through the \ce{N-CH2N} complex, or via direct abstraction. We find the addition reaction forming the \ce{N-CH2N} complex to have a large barrier at the BH/d level of theory (82 kJ mol$^{-1}$).

Furthermore, we find the addition reaction forming \ce{^3H2CNN} to efficiently decay into the \ce{N2} and \ce{^3CH2} products. \ce{^3H2CNN} isomerization barriers proceeding to \ce{HCN + NH} decay are too large to consider this pathway.

At the BH/d and CC/t levels of theory, we find the direct abstraction reaction to have barriers of 15.4 and 23.0 kJ mol$^{-1}$, respectively. On the other hand, at the B3LYP/aug-cc-pVDZ and $\omega$B/d levels of theory, we find the abstraction reaction to be barrierless and nearly barrierless (E$_0$ = 1.9 kJ mol$^{-1}$), respectively.

Including the barriers, the rate coefficients for \ce{H2CN + ^4N -> HCN + NH} at the BH/d and CC/t levels of theory are 9.5$\times$10$^{-15}$ and 7.1$\times$10$^{-14}$ cm$^{3}$s$^{-1}$, respectively. These are several orders of magnitude lower than the experimentally measured value. 

We find it likely that BH/d and CC/t calculate barriers when there should not be any. When removing the barriers from the calculation, the rate coefficients at the BH/d and CC/t levels of theory are 4.7$\times$10$^{-12}$ and 5.1$\times$10$^{-12}$ cm$^{3}$s$^{-1}$, respectively. This reduces the discrepancy between experiment and calculation to factors of 9.

The rate coefficients at the $\omega$B/d and B3LYP/aug-cc-pVDZ levels of theory are 3.4 and 9.4$\times$10$^{-12}$ cm$^{3}$s$^{-1}$, which are factors of 13 and 5 smaller than the experimental value. Given all these discrepancies, we recommend further experimental and theoretical analyses be carried out for this reaction.

\subsubsection*{Case Study 2: \ce{CN + N -> CN2$_{(\nu)}$* -> N2 + C}}

Experimental measurements of this reaction near 298 K range from 1.0 to 1.2$\times$10$^{-10}$ cm$^{3}$s$^{-1}$ \citep{Reference2003,Reference2004}. \citet{Reference451} reviewed both high- and low-temperature measurements and recommended a value of $k$ = 3.0$\times$10$^{-10}$ cm$^{3}$s$^{-1}$.

The mechanism of the reaction is not previously well understood, however authors have suggested this reaction may pass through the \ce{CN2*} intermediate \citep{Reference2004,Reference451}.
Our theoretical simulations suggest this is indeed the correct mechanism.

We calculate the rate coefficient of \ce{CN + N -> CN2} at 298 K using the BH/d level of theory to be 3.4$\times$10$^{-11}$ cm$^{3}$s$^{-1}$. We find the decay of \ce{CN2$_{(\nu=0)}$} into \ce{N2 + C} to be slow ($\sim$10$^{-41}$ s$^{-1}$). This suggests that this reaction likely proceeds through vibrationally excited \ce{CN2}, as is to be expected when two reactants combine to form a single product \citep{Vallance_Book}.

Our calculated rate coefficient for \ce{CN + N -> CN2$_{(\nu)}$* -> N2 + C} (4.3$\times$10$^{-11}$ cm$^{3}$s$^{-1}$) is only a factor of 2 smaller than the experimentally measured values.

\subsubsection*{Case Study 3: \ce{CN + CH4 -> HCN + CH3}}

The experimental rate coefficient for this reaction at 298 K ranges from 5.6$\times$10$^{-13}$ to 1.1$\times$10$^{-12}$ cm$^{3}$s$^{-1}$ \citep{Reference2006,Reference2007,Reference2008,Reference2009,Reference2010,Reference2011,Reference2012,Reference2013}.

The experimental barrier for this reaction is about 8.3 kJ mol$^{-1}$ \citep{Reference2051}. This matches well with the MP4 theoretical barriers calculated by \citet{Reference2005} which range from 7.5--8.8 kJ mol$^{-1}$. We calculate a barrier for this reaction at the CC/t level of theory to be 6.7 kJ mol$^{-1}$.

We do not calculate a barrier for this reaction at the BH/d or $\omega$B/d levels of theory. We calculate a barrierless rate coefficients at 298 K at to be 2.2 and 3.7 $\times$10$^{-11}$ cm$^{3}$s$^{-1}$, respectively, which are over an order of magnitude greater than the nearest experimental value. This disagreement with experiment is due to the lack of barriers calculated at these levels of theory. For this reason, we introduce the experimental barrier of 8.3 kJ mol$^{-1}$ \citep{Reference2051} to our BH/d calculation. This produces a rate coefficient at 298 K of 7.7$\times$10$^{-13}$ cm$^{3}$s$^{-1}$, which is within the experimental range.

\subsubsection*{Case Study 4: \ce{CN + H2 -> HCN + H}}

Experiments measure the rate coefficient for this reaction to be between 1.2 and 4.9$\times$10$^{-14}$ cm$^{3}$s$^{-1}$ \citep{Reference2012,Reference2013,Reference2014,Reference2015,Reference2016,Reference2017,Reference2018,Reference2019,Reference2020}.

Experiments generally agree on a barrier for this reaction of $\sim$16.7 kJ mol$^{-1}$ \citep{Reference2052}. However, an {\it ab initio} theoretical study suggests a barrier of 13.4 kJ mol$^{-1}$ provides a much better agreement between theory and experiment \citep{Reference2053}.

We calculate no barrier for this reaction at the BH/d level of theory. The barrierless rate coefficient we calculate at 298 K is 8.7$\times$10$^{-13}$ cm$^{3}$s$^{-1}$, which is a factor of 18 larger than the nearest experimental value. This discrepancy is due to the lack of barrier in our calculation when one should be present.

Similarly to \citet{Reference2053}, when including a barrier of $\sim$16.7 kJ mol$^{-1}$, the calculated rate coefficient at 298 K is too low with respect to experimental values. Therefore, we include an experimental barrier of 13.4 kJ mol$^{-1}$ in our calculation and obtain a rate coefficient of $k$(298 K) = 3.9$\times$10$^{-15}$ cm$^{3}$s$^{-1}$. This value is only a factor of 3 smaller than the nearest experimental value. 

\subsubsection*{Case Study 5: \ce{CH4 + ^1CH2 -> C2H6$_{(\nu)}$* -> CH3 + CH3}}

\ce{^1CH2} insterts into the C-H bond of \ce{CH4} to produce the \ce{C2H6} intermediate, and subsequently dissociates into \ce{CH3 + CH3} \citep{Reference514,Reference515,Reference509}.

We calculate the decay of \ce{C2H6} in the ground vibrational state into \ce{CH3 + CH3} to be slow ($<$ 10$^{-37}$ s$^{-1}$), which suggests this reaction proceeds through an excited vibrational state.

In previous work we only included the reaction \ce{CH4 + ^1CH2 -> C2H6} in the network \citep{Reference598}; However this is only valid in high atmospheric pressures. We therefore include calculations of both the high pressure (\ce{CH4 + ^1CH2 + M -> C2H6 + M} and low pressure reactions (\ce{CH4 + ^1CH2 -> C2H6$_{(\nu)}$* -> CH3 + CH3}), with BH/d rate coefficients at 298 K of 7.2$\times$10$^{-24}$ cm$^{6}$s$^{-1}$ and 2.1$\times$10$^{-11}$ cm$^{3}$s$^{-1}$, respectively.

\subsubsection*{Case Study 6: \ce{CH4 + CH -> CH4-CH* -> C2H5* -> C2H4 + H}}

Experimentally measured rate coefficients for this reaction range from $k$(298) = 0.02--3$\times$10$^{-10}$ cm$^{3}$s$^{-1}$ \citep{Reference521,Reference562,Reference561,Reference563,1997AA...323..644C,Reference566,Reference520,Reference519,Reference523}. Experiment suggests this reaction proceeds without a barrier \citep{Reference561}.

Previous theoretical studies find this reaction proceeds through three steps, first forming a \ce{CH4-CH} complex, then \ce{C2H5}, and finally, decay into \ce{C2H4 + H} \citep{Reference2063,Reference565}. At the CCSD(T)-F12/CBS//B3LYP/6-311G(d,p) + ZPE(B3LYP/6-311G(d,p)) level of theory, \citet{Reference2063} found the first step of this reaction to be barrierless, and the second step to have a barrier 2.3 kJ mol$^{-1}$ lower than the reactants. At the UMP2/6-31G(d,p) and UMP4/6-311++(2d,p)//UMP2/6-31G(d,p) levels of theory, \citet{Reference565} found similar results, with a barrierless first step, and second step barriers 0.3 and 0.2 kJ mol$^{-1}$ below the reactants, respectively.

At the BH/d level of theory, we also find the first step to be barrierless; However, we find the \ce{CH4-CH* -> C2H5*} step to have a barrier 11.5 kJ mol$^{-1}$ above the reactants. This is similar to our result at the CC/t level of theory, where we find this step to have a barrier 7.7 kJ mol$^{-1}$ above the reactants.

At the $\omega$B/d and B3LYP/aug-cc-pvDZ levels of theory, we find the \ce{CH4-CH* -> C2H5*} step to be barrierless. 

Because theory and experiment suggest the \ce{CH4-CH* -> C2H5*} step is not rate-limiting, we remove the barrier from this calculation. This adjustment leads to a BH/d calculated rate coefficient of 1.8$\times$10$^{-10}$ cm$^{3}$s$^{-1}$, which is within the experimental range.





\subsubsection*{Case Study 7: \ce{^3CH2 + H <-> CH3$_{(\nu)}$* <-> CH + H2}}

Reactions of \ce{^3CH2 + H} and \ce{CH + H2} are suggested to produce vibrationally excited \ce{CH3$_{(\nu)}$} \citep{Reference503}. This is what is to be expected when two reactants combine to form a single product \citep{Vallance_Book}. In high atmospheric pressure, \ce{CH3$_{(\nu)}$} collisionally deexcites \ce{CH3$_{(\nu)}$ + M -> CH3 + M} \citep{Reference503,Reference516}. In low atmospheric pressure, \ce{CH3$_{(\nu)}$} dissociates into \ce{CH + H2} approximately 80$\%$ of the time, and \ce{^3CH2 + H} approximately 20$\%$ of the time \citep{Reference2040}.

In previous work, we only included reactions \ce{^3CH2 + H -> CH3} and \ce{CH + H2 -> CH3} in the network \citep{Reference598}. These are only valid for high atmospheric pressures; Therefore, we modify the network to include the three-body reactions, 

\begin{equation*}
\ce{^3CH2 + H + M -> CH3 + M},
\end{equation*}
\begin{equation*}
\ce{CH + H2 + M -> CH3 + M}.
\end{equation*}

We calculate these rate coefficients at the BH/d level of theory to be $k$ =  4.8 and 7.8$\times$10$^{-29}$ cm$^{6}$s$^{-1}$, respectively. 

We also include the two-body reactions,
\begin{equation*}
\ce{^3CH2 + H -> CH3$_{(\nu)}$* -> CH + H2}, 
\end{equation*}
\begin{equation*}
\ce{CH + H2 -> CH3$_{(\nu)}$* -> ^3CH2 + H},
\end{equation*}
with adjusted calculated rate coefficients $k$ = 3.7$\times$10$^{-10}$ and 5.4$\times$10$^{-11}$ cm$^{3}$s$^{-1}$, respectively.

The rate coefficients for \ce{^3CH2 + H -> CH3$_{(\nu)}$* -> CH + H2} and \ce{CH + H2 -> CH3$_{(\nu)}$* -> ^3CH2 + H} are reduced by 80$\%$ and 20$\%$ based on the \ce{CH3$_{(\nu)}$} dissociation probabilities calculated by \citet{Reference2040}.

The same treatment is applied to the reaction of excited state \ce{^1CH2} with \ce{H}. We calculate the reaction rate coefficient for \ce{^1CH2 + H + M -> CH3 + M} to be 3.7$\times$10$^{-28}$ cm$^{6}$s$^{-1}$.
We also adjust our previously calculated reaction rate coefficient for \ce{^1CH2 + H -> CH3} \citep{Reference598} to dissociate along the channels \ce{CH + H2} and \ce{^3CH2 + H} with the same branching ratios as above. These new rate coefficients are 1.8$\times$10$^{-10}$ and 4.6$\times$10$^{-11}$ cm$^{3}$s$^{-1}$, respectively.

\subsubsection*{Case Study 8: \ce{^1CH2 + H2 -> CH4$_{(\nu)}$* -> CH3 + H}}

The reaction of \ce{^1CH2 + H2} proceeds through an excited methane molecule before most commonly decaying into \ce{CH3 + H} \citep{Reference514,Reference510}.

In previous work, we only included the reaction \ce{^1CH2 + H2 -> CH4} in the network \citep{Reference598}. This is only valid for high pressure environments; Therefore we now include the three-body reaction \ce{^1CH2 + H2 + M -> CH4 + M}, and we apply our previously calculated rate coefficient to the two-body reaction \ce{^1CH2 + H2 -> CH4$_{(\nu)}$* -> CH3 + H}, by assuming vibrational decay into \ce{CH3 + H}. These reactions have calculated rate coefficients $k$(298 K) = 1.4$\times$10$^{-27}$ cm$^{6}$s$^{-1}$, and 1.0$\times$10$^{-10}$ cm$^{3}$s$^{-1}$, at the BH/d level of theory, respectively.

\subsubsection*{Case Study 9: \ce{CH + H -> ^3CH2$_{(\nu)}$* -> C + H2}}

The reaction of \ce{CH + H} proceeds through \ce{^3CH2} before decaying into \ce{C + H2} \citep{Reference2041}. It is expected that the intermediate would be vibrationally exited \citep{Vallance_Book}.

In previous work, we only included \ce{CH + H -> ^3CH2} in the network \citep{Reference598}; However, this is only valid in the high-pressure limit. We now include both the three-body reaction \ce{CH + H + M -> ^3CH2 + M} and the low-pressure vibrational decay reaction 
\ce{CH + H -> ^3CH2$_{(\nu)}$* -> C + H2} with calculated rate coefficients $k$(298) = 2.0$\times$10$^{-28}$ cm$^{6}$s$^{-1}$, and 6.9$\times$10$^{-10}$ cm$^{3}$s$^{-1}$, at the BH/d level of theory, respectively.

We also adjust our previously calculated reaction \ce{CH + H -> ^1CH2} to be a three-body reaction with a low pressure rate coefficient of 4.4$\times$10$^{-31}$ cm$^{6}$s$^{-1}$. We assume that in low pressures the \ce{^1CH2$_{(\nu)}$} molecule vibrationally decays back into \ce{CH + H}.

\section*{Quantum Chemistry Data}

\setlength\LTcapwidth{\textwidth}
\begin{longtable*}{cclcccccc}
\caption{Quantum Chemistry Data at the BHandHLYP/aug-cc-pVDZ level of theory. Cartesian coordinates are in angstroms. Energies are in kJ mol$^{-1}$. E$_e$ is the electronic energy, ZPE is the zero point energy, and q$_x$ are the partition functions. \label{quantumdata}} \\
Reac. & Species & Geometry (Atom, X, Y, Z) & E$_e$ + ZPE & E$_e$ + Gibbs & q$_t$/V(m$^{-3}$) & q$_e$ & q$_v$ & q$_r$ \\ \hline \\[-2mm]
1 & HCN & N, 0.00000, 0.00000, 0.64811 & -245122.398339 & -245173.207015 & 1.36E+32 & 1 & 1.05E+00	& 1.38E+02 \\
 & & C, 0.00000, 0.00000, -0.49567 & & & & & & \\
& & H, 0.00000, 0.00000, -1.56273 & & & & & & \\
 & CH & C, 0.00000, 0.00000, 0.16040 & -100985.100094 & -101029.197992 & 4.54E+31 &	2 &	1.00E+00	& 1.44E+01 \\
& & H, 0.00000, 0.00000, -0.96239 & & & & & & \\
\hline
\end{longtable*}

\setlength\LTcapwidth{\textwidth}
\begin{longtable*}{cclcccccc}
\caption{Quantum Chemistry Data at the $\omega$B97XD/aug-cc-pVDZ level of theory. Cartesian coordinates are in angstroms. Energies are in kJ mol$^{-1}$. \label{quantumdata2}} \\
Reac. & Species & Geometry (Atom, X, Y, Z) & E$_e$ + ZPE & E$_e$ + Gibbs & q$_t$/V (m$^{-3}$) & q$_e$ & q$_v$ & q$_r$ \\ \hline \\[-2mm]
22 & \ce{CH3} & C, 0.00000, -0.00000, 0.00000 & -104483.836143 & -104535.821043 & 5.63E+31 &	2 &	1.08E+00 &	2.58E+02 \\
& & H, -0.88473, -0.63289, 0.00000 & & & & & & \\
& & H, -0.10575, 1.08262, 0.00000 & & & & & & \\
& & H, 0.99047, -0.44973, 0.00000 & & & & & & \\
 & H & H, 0.00000, 0.00000, 0.00000 & -1318.707191 & -1318.707191 & 9.79E+29	& 2 &	1.00E+00 & 	1.00E+00 \\
\hline
\end{longtable*}

\setlength\LTcapwidth{\textwidth}
\begin{longtable*}{cclcccccc}
\caption{Quantum Chemistry Data at the CCSD/aug-cc-pVTZ level of theory. Cartesian coordinates are in angstroms. Energies are in kJ mol$^{-1}$. \label{quantumdata3}} \\
Reac. & Species & Geometry (Atom, X, Y, Z) & E$_e$ + ZPE & E$_e$ + Gibbs & q$_t$/V (m$^{-3}$) & q$_e$ & q$_v$ & q$_r$ \\ \hline \\[-2mm]
22 & \ce{CH3} & C, 0.00000, 0.00000, 0.00000 & -104308.024787 & -104359.991308 & 5.63E+31 &	2 &	1.10E+00 &	2.52E+02 \\
& & H, -0.90220, 0.59027, 0.00000 & & & & & & \\
& & H,  0.96228, 0.48619, 0.00000 & & & & & & \\
& & H, -0.06009, -1.07646, 0.00000 & & & & & & \\
& H &  H, 0.00000, 0.00000, 0.00000 & -1312.280561 & -1312.280561 & 9.79E+29 &	2 & 1.00E+00 & 	1.00E+00 \\
\hline
\end{longtable*}

\setlength\LTcapwidth{\textwidth}
\begin{longtable*}{cclcccccc}
\caption{Quantum Chemistry Data at the BHandHLYP/aug-cc-pVTZ level of theory. Cartesian coordinates are in angstroms. Energies are in kJ mol$^{-1}$. \label{quantumdata3}} \\
Reac. & Species & Geometry (Atom, X, Y, Z) & E$_e$ + ZPE & E$_e$ + Gibbs & q$_t$/V (m$^{-3}$) & q$_e$ & q$_v$ & q$_r$ \\ \hline \\[-2mm]
22 & \ce{CH3} & C, 0.00000, 0.00000, 0.00001 & -104491.954189 & -104543.826193 & 5.63E+31 &	2 &	1.08E+00 &	2.46E+02 \\
& & H, -0.81893, -0.68980, -0.00001 & & & & & & \\
& & H, -0.18792, 1.05411, -0.00001 & & & & & & \\
& & H, 1.00685, -0.36432, -0.00001 & & & & & & \\
& H &  H, 0.00000, 0.00000, 0.00000 & -1309.084906 & -1309.084906 & 9.79E+29 &	2 & 1.00E+00 & 	1.00E+00 \\
\hline
\end{longtable*}

\end{document}